\documentclass[aps,twocolumn,floatfix,superscriptaddress,pra]{revtex4-1}
\usepackage[latin9]{inputenc}
\usepackage{amsmath}
\usepackage{amssymb}
\usepackage{graphicx}
\usepackage{esint}
\usepackage{epstopdf}
\usepackage{braket}
\usepackage{color}
\usepackage{hyperref}
\usepackage{tikz-feynman}
\usepackage{bibunits}
\usepackage{mathtools}

\defaultbibliography{Entanglement.bib}
\defaultbibliographystyle{apsrev4-1}

\makeatletter
\@ifundefined{textcolor}{}
{
 \definecolor{BLACK}{gray}{0}
 \definecolor{WHITE}{gray}{1}
 \definecolor{RED}{rgb}{1,0,0}
 \definecolor{GREEN}{rgb}{0,1,0}
 \definecolor{BLUE}{rgb}{0,0,1}
 \definecolor{CYAN}{cmyk}{1,0,0,0}
 \definecolor{MAGENTA}{cmyk}{0,1,0,0}
 \definecolor{YELLOW}{cmyk}{0,0,1,0}
}

\makeatother

\begin{document}

\title{Entanglement and quantum tomography with top quarks at the LHC}

\author{Yoav Afik}
\email{yoavafik@gmail.com}
\affiliation{Experimental Physics Department, CERN, 1211 Geneva, Switzerland}

\author{Juan Ram\'on Mu\~noz de Nova}
\email{jrmnova@fis.ucm.es}
\affiliation{Departamento de F\'isica de Materiales, Universidad Complutense de Madrid, E-28040 Madrid, Spain}

\begin{abstract}
Entanglement is a central subject in quantum mechanics. Due to its genuine relativistic behavior and fundamental nature, high-energy colliders are attractive systems for the experimental study of fundamental aspects of quantum mechanics. We propose the detection of entanglement between the spins of top-antitop-quark pairs at the LHC, representing the first proposal of entanglement detection in a pair of quarks, and also the entanglement observation at the highest energy scale so far. We show that entanglement can be observed by direct measurement of the angular separation between the leptons arising from the decay of the top-antitop pair. The detection can be achieved with high statistical significance, using the current data recorded during Run 2 at the LHC. In addition, we develop a simple protocol for the quantum tomography of the top-antitop pair. This experimental technique reconstructs the quantum state of the system, providing a new experimental tool to test theoretical predictions. Our work explicitly implements canonical experimental techniques in quantum information in a two-qubit high-energy system, paving the way to use high-energy colliders to also study quantum information aspects.
\end{abstract}

\date{\today}

\maketitle
\section{Introduction}
Entanglement is one of the most genuine features of quantum mechanics~\cite{Einstein:1935rr,Schrodinger1935,Bell:1964kc}. It has been observed in systems as different as photons~\cite{Aspect1982}, atoms~\cite{Hagley1997}, superconductors~\cite{Steffen2006}, mesons~\cite{Go:2007ww}, neutrinos~\cite{Formaggio2016}, analog Hawking radiation~\cite{Steinhauer2016}, nitrogen-vacancy centers in diamond~\cite{Pfaff2013} and even macroscopic diamond itself~\cite{Lee2011}. Recently, loophole-free violations of Bell's inequality were measured for the first time~\cite{Hensen2015,Giustina2015}. Apart from its fundamental interest, entanglement is a key element in metrology, teleportation, quantum information and quantum computation~\cite{Giovannetti2004,Bouwmeester1997,Nielsen2000}.

The standard model (SM) of particle physics is a quantum field theory, based on special relativity and quantum mechanics. Therefore, it allows to test fundamental properties of quantum mechanics. For instance, entanglement has been studied in the context of particle physics~\cite{Bramon2002,Shi2004,Kayser2010,Alba2017} and is currently a hot topic of research in the field~\cite{Shi2020,Tu2020,Ramos2020,Aoud2020,Fabbrichesi2021}.

The most simple model in quantum mechanics is a qubit, a two-level quantum system. The canonical example of entanglement is provided by two qubits, as it is the case of two spin-1/2 particles, where the entanglement is characterized by their spin correlations. At the LHC, the spin correlations between two particles can be extracted from the kinematical distributions of the products of their decay. Top quarks fulfill the requirements for spin correlations measurements~\cite{Tanabashi:2018oca}: they have a lifetime ($\sim 10^{-25} \textrm{s}$) which is shorter than the time scale for hadronisation ($\sim 10^{-23} \textrm{s}$) and for spin decorrelation ($\sim 10^{-21} \textrm{s}$).

The framework of spin correlations between top quarks, originally developed in Refs.~\cite{Kane1992,Bernreuther1994,Bernreuther1998,Bernreuther2001,Bernreuther2004}, is an intensive field of study by itself~\cite{Mahlon1996,Parke1996,Uwer2005,Baumgart2013,Bernreuther2015}. Measurements of the top-antitop ($t\bar{t}$) spin correlations have been provided by the D0 and CDF collaborations at the Tevatron with proton-antiproton ($p\bar{p}$) interactions~\cite{Abazov:2011ka,Abazov:2011qu,Abazov:2011gi,Abazov:2015psg,Aaltonen:2010nz}, and by the ATLAS and CMS collaborations at the LHC with proton-proton ($pp$) interactions~\cite{ATLAS:2012ao,Aad:2014pwa,Aad:2014mfk,Aad:2015bfa,Aaboud:2016bit,Aaboud:2019hwz,Chatrchyan:2013wua,Khachatryan:2015tzo,Khachatryan:2016xws,Sirunyan:2019lnl}. However, so far, no link between top spin correlations and entanglement has been discussed at the literature. 

We present here the first study of spin entanglement between top quarks, and show that the entanglement between the spins of a $t\bar{t}$ pair can be detected at the LHC, which represents the first proposal of detection of entanglement between a pair of quarks. The entanglement can be directly measured from the angular separation of the $t \bar{t}$ lepton decay products, a measurable observable in experiments~\cite{Sirunyan:2019lnl}, with high statistical significance and using currently recorded data by Run 2 at the LHC. However, due to the nature of $t \bar{t}$ production, the observation is far from trivial, and it can only be performed within a restricted region of phase space with the help of a dedicated analysis. In addition to the intrinsic value of testing entanglement for the first time in a new scenario, our work suggests a new perspective within the well-established subject of top spin correlations.

Furthermore, we develop a simple experimental protocol for the quantum tomography (i.e., the reconstruction of the quantum state) of the $t\bar{t}$ pair, which involves similar observables to those in current measurements of top spin correlations~\cite{Sirunyan:2019lnl}, and extends previous theoretical work on quantum tomography in high-energy physics \cite{Martens2018}. The quantum tomography of the $t \bar{t}$ pair goes beyond the entanglement detection and provides a new experimental platform to test theoretical predictions since it gives full access to the quantum state, which contains all the information about a system. For instance, one can try to measure possible effects arising from new physics beyond the standard model that modify the quantum state of the $t \bar{t}$ pair.

Our proposal implements canonical quantum information techniques in a two-qubit system at the LHC. Thus, although top quarks are not useful for information transmission due to their short lifetime, the present work paves the way for the study of quantum information aspects at high-energy colliders. Due to its genuine relativistic behavior and fundamental nature, the high-energy physics environment is particularly interesting for the study of quantum information and other elemental topics in quantum mechanics. For instance, relativistic effects are expected to play a key role in quantum information and, especially, in the nature of entanglement itself~\cite{Gingrich2002,Peres2004,Friis2010,Friis2013,Giacomini2019}. Indeed, our work proposes the highest-energy entanglement detection and quantum tomography protocol ever made.

The paper is arranged as follows. Section~\ref{sec:general} discusses in detail the general framework upon we build the results of this work. Section~\ref{sec:entanglement} studies entanglement in $t\bar{t}$ production. Section~\ref{sec:totalquantumstate} characterizes, with the help of the previous results, the presence of entanglement in the total quantum state of a $t\bar{t}$ pair produced at the LHC. Section~\ref{sec:QuantumTomography} proposes an experimental scheme for the implementation of the quantum tomography of the $t\bar{t}$ pair. Section~\ref{sec:experimental} assesses the experimental detection of entanglement and its statistical significance at the LHC. Section~\ref{sec:conclusions} provides conclusions and outlook. Technical details are presented in the Appendix.

\section{General framework}\label{sec:general}

\subsection{Spin-$1/2$ bipartite systems}\label{subsec:Hilbert}

A quantum state is described in general by a density matrix $\rho$, a nonnegative operator in a Hilbert space satisfying $\rm{tr}(\rho)=1$. The expectation value of a certain observable $O$ in that quantum state is obtained as $\braket{O}=\rm{tr}(O\rho)$. For a bipartite Hilbert space $\mathcal{H}$ formed by the direct product of two sub-systems $a,b$, $\mathcal{H}=\mathcal{H}_a\otimes\mathcal{H}_b$, a quantum state is said to be separable if it can be written as
\begin{equation}\label{eq:Separability}
\rho=\sum_n p_n \rho^{a}_{n}\otimes\rho^{b}_{n},~\sum_n p_n=1,~p_n\geq0 ,
\end{equation}
where $\rho^{a,b}_n$ are quantum states in the sub-systems $a,b$. Any classically correlated state in $\mathcal{H}$ can be put in this form~\cite{Werner1989}. A state that is non-separable is called entangled and hence it is a non-classical state.

The canonical case of a bipartite system is the Hilbert space formed by two qubits, of dimension $2\times 2$, where the most general form for a density matrix $\rho$ describing a quantum state in $\mathcal{H}$ is
\begin{equation}\label{eq:GeneralBipartiteStateRotations}
\rho=\frac{I_4+\sum_{i}\left(B^{+}_{i}\sigma^i\otimes I_2+B^{-}_{i}I_2\otimes\sigma^i \right)+\sum_{i,j}C_{ij}\sigma^{i}\otimes\sigma^{j}}{4}
\end{equation}
with $i,j=1,2,3$, $I_n$ the $n\times n$ identity matrix and $\sigma^i$ the corresponding Pauli matrices. Thus, a general quantum state in a Hilbert space of dimension $2\times 2$ is determined by $15$ parameters, $B^{\pm}_{i},C_{ij}$. Specifically, in the case of two spin-1/2 particles, the vectors $\mathbf{B}^{\pm}$ characterize the individual spin polarization of each particle, $B^{+}_{i}=\braket{\sigma^i\otimes I_2},~B^{-}_{i}=\braket{I_2\otimes\sigma^i}$, while the correlation matrix $\mathbf{C}$ characterizes the spin correlations between the particles, $C_{ij}=\braket{\sigma^{i}\otimes\sigma^{j}}$.  As a result, by measuring all these $15$ expectation values, $\braket{\sigma^i\otimes I_2}$, $\braket{I_2\otimes\sigma^i}$, $\braket{\sigma^{i}\otimes\sigma^{j}}$, one can experimentally reconstruct the quantum state of the two qubits. This simple idea is the basis of quantum tomography.

An important criterion for signaling entanglement in bipartite systems is the Peres-Horodecki criterion~\cite{Peres1996,Horodecki1997}. The criterion is simply based on the observation that if $\rho$ is separable, the state resulting from taking the partial transpose in, for instance, the second subsystem,
\begin{equation}\label{eq:Separability}
\rho^{\rm{T_2}}=\sum_n p_n \rho^{a}_{n}\otimes\left(\rho^{b}_{n}\right)^{\rm{T}},
\end{equation}
should be also a physical state, i.e., a nonnegative operator with unit trace. Therefore, if $\rho^{\rm{T_2}}$ is not nonnegative, the state is entangled. The Peres-Horodecki criterion is also a necessary condition for entanglement in bipartite systems of dimension $2\times 2$.

A more quantitative measurement of the degree of entanglement is provided by the concurrence $C[\rho]$, which is a function of the quantum state that is related to the entanglement of formation~\cite{Wooter1998}. Specifically,
\begin{equation}\label{eq:Concurrence}
C[\rho]\equiv \max(0,\lambda_1-\lambda_2-\lambda_3-\lambda_4)
\end{equation}
where $\lambda_i$ are the eigenvalues, ordered in decreasing magnitude, of the matrix $\mathcal{C}(\rho)=\sqrt{\sqrt{\rho}\tilde{\rho}\sqrt{\rho}}$, with $\tilde{\rho}=(\sigma_2\otimes\sigma_2)~\rho^*~(\sigma_2\otimes\sigma_2)$ and $\rho^*$ the complex conjugate of the density matrix in the usual spin basis of $\sigma_3$. The concurrence satisfies $0\leq C[\rho]\leq 1$, with a quantum state being entangled if and only if $C[\rho]>0$. Therefore, states satisfying $C[\rho]=1$ are maximally entangled. We refer the reader to Appendix~\ref{app:criteria} for the derivation of some useful results, based on these entanglement criteria, used throughout this work.

\subsection{$t\bar{t}$ production}\label{subsec:ttgeneral}

An example of a two-qubit system is provided by a pair of quarks, which are spin-1/2 particles, as it is the case of a $t\bar{t}$ pair. At the LHC, a $t\bar{t}$ pair arises from $pp$ collisions at high energies. A proton consists of quarks (spin-1/2 fermions) and gluons (massless spin-1 bosons), which are indistinctively denoted as partons~\cite{Feynman:1969wa,Bjorken:1969ja}.
The composition of the proton is modeled by the so-called parton distribution function (PDF), which determines the density of each parton in the proton by the momentum transfer.

\begin{figure}[tb!]
\centering
\includegraphics[width=\columnwidth]{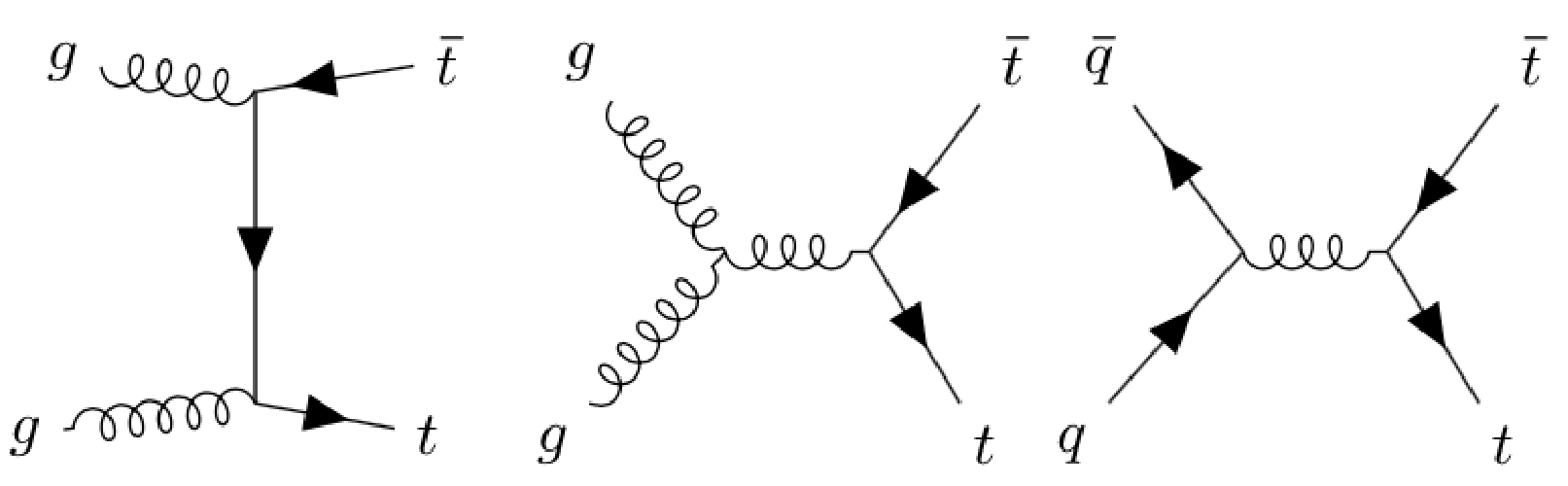}
     \caption{Representative Feynman diagrams for a $t \overline{t}$ production via the SM.}
     \label{fig:Feynman}
\end{figure}

Interactions between these partons through quantum chromodynamics (QCD) give rise to a $t \bar{t}$ pair. For instance, a $t\bar{t}$ pair can arise from the interaction between a light quark and antiquark ($q \bar{q}$), or between a pair of gluons ($gg$),
\begin{eqnarray}\label{eq:partonreacitons}
    q+\bar{q}&\rightarrow& t+\bar{t} ,\\
\nonumber g+g&\rightarrow& t+\bar{t} .
\end{eqnarray}
Representative Feynman diagrams for these processes are presented in Fig.~\ref{fig:Feynman}.

Kinematically, the production of a $t\bar{t}$ pair is described by the invariant mass $M_{t\bar{t}}$ and the top direction $\hat{k}$ in the center-of-mass (CM) frame. Specifically, in this frame the top and antitop relativistic momenta are $k^{\mu}_t=(k_t^0,\mathbf{k}),
k^{\mu}_{\bar{t}}=(k_{\bar{t}}^0,-\mathbf{k})$, satisfying the invariant dispersion relation $k^2_t\equiv k^{\mu}_t k_{{\mu}t}=m^2_t$, and similar for the antitop $k^2_{\bar{t}}=k^2_t=m^2_t$. The invariant mass is defined from these momenta as
\begin{equation}
    M^2_{t\bar{t}}\equiv s_{t\bar{t}}=(k_t+k_{\bar{t}})^2,
\end{equation}
with $s_{t\bar{t}}$ the usual Mandelstam variable. In the CM frame, this gives $M^2_{t\bar{t}}=4\left(k_t^0\right)^2=4(m^2_t +\mathbf{k}^2)$. By relating the top momentum to its velocity $\beta$ by $|\mathbf{k}|=m_t\beta/\sqrt{1-\beta^2}$, we get
\begin{equation}
    \beta=\sqrt{1-4m^2_t/M^2_{t\bar{t}}} ,
\end{equation}
from where we immediately see that threshold production ($\beta=0$) corresponds to $M_{t\bar{t}}=2m_t \approx 346~\textrm{GeV}$, the minimum energy possible for a $t\bar{t}$ pair.

While the kinematics of the $t\bar{t}$ pair are determined by the variables $(M_{t\bar{t}},\hat{k})$, their spins for a fixed production process are characterized by the so-called production spin density matrix $R(M_{t\bar{t}},\hat{k})$~\cite{Bernreuther1994}, whose most general form is similar to that of Eq. (\ref{eq:GeneralBipartiteStateRotations}),
\begin{equation}\label{eq:GeneralDensityMatrix}
R=\tilde{A}I_4+\sum_{i}\left(\tilde{B}^{+}_{i}\sigma^i\otimes I_2+\tilde{B}^{-}_{i}I_2\otimes\sigma^i \right)+\sum_{i,j}\tilde{C}_{ij}\sigma^{i}\otimes\sigma^{j}
\end{equation}
where the first/second spin subspace corresponds to the top/antitop, respectively. We note that the production spin density matrix is characterized by $16$ parameters, $\tilde{A},\tilde{B}^{\pm}_{i},\tilde{C}_{ij}$, one more than in Eq. (\ref{eq:GeneralBipartiteStateRotations}). This is because the matrix $R$ is not properly normalized since $\mathrm{tr}(R)=4\tilde{A}$, with $\tilde{A}$ determining the differential cross section for $t\bar{t}$ production at fixed energy and top direction,
\begin{equation}\label{eq:CrossSectionDifferential}
\frac{\mathrm{d}\sigma}{\mathrm{d}\Omega\mathrm{d}M_{t\bar{t}}}=\frac{\alpha^2_s\beta}{M^2_{t\bar{t}}}\tilde{A}(M_{t\bar{t}},\hat{k})
\end{equation}
$\Omega$ being the solid angle associated with $\hat{k}$ and $\alpha_s\approx 0.118$ the strong coupling constant.

The proper spin density matrix $\rho(M_{t\bar{t}},\hat{k})$ of Eq. (\ref{eq:GeneralBipartiteStateRotations}) and the actual spin polarizations $B^{\pm}_{i}$ and spin correlations $C_{ij}$ of the $t\bar{t}$ pair are
simply computed by normalizing $R$,
\begin{equation}\label{eq:PhysicalPolCor}
\rho=\frac{R}{\mathrm{tr}(R)}=\frac{R}{4\tilde{A}},~B^{\pm}_{i}=\frac{\tilde{B}^{\pm}_{i}}{\tilde{A}},~C_{ij}=\frac{\tilde{C}_{ij}}{\tilde{A}} .
\end{equation}

With the help of the production spin density matrix, the expectation value of any observable in a selected region $\Pi$ of the phase space $(M_{t\bar{t}},\hat{k})$ is obtained by integration as~\cite{Bernreuther1998,Uwer2005}
\begin{equation}\label{eq:ExpectationValueR}
\braket{O}=\frac{\int_\Pi\mathrm{d}\Omega\mathrm{d}M_{t\bar{t}}~\frac{\alpha^2_s\beta}{M^2_{t\bar{t}}}\mathrm{tr}(OR)}
{\int_\Pi\mathrm{d}\Omega\mathrm{d}M_{t\bar{t}}~\frac{\alpha^2_s\beta}{M^2_{t\bar{t}}}\mathrm{tr}(R)} .
\end{equation}
In terms of the proper spin density matrices $\rho(M_{t\bar{t}},\hat{k})$, the above equation simply reads
\begin{equation}\label{eq:Expectationvalue}
\braket{O}=\int_\Pi\mathrm{d}M_{t\bar{t}}\mathrm{d}\Omega~p(M_{t\bar{t}},\hat{k})\braket{O}_{\rho} ,
\end{equation}
with $\braket{O}_{\rho}\equiv\mathrm{tr}[O\rho(M_{t\bar{t}},\hat{k})]$ and
\begin{eqnarray}\label{eq:individualquantumstates}
\nonumber p(M_{t\bar{t}},\hat{k})&=&\frac{1}{\sigma_\Pi}\frac{\mathrm{d}\sigma}{\mathrm{d}\Omega\mathrm{d}M_{t\bar{t}}}\\
\sigma_\Pi&\equiv&\int_\Pi\mathrm{d}\Omega\mathrm{d}M_{t\bar{t}}~\frac{\mathrm{d}\sigma}{\mathrm{d}\Omega\mathrm{d}M_{t\bar{t}}} .
\end{eqnarray}

The expectation value in Eq. (\ref{eq:Expectationvalue}) can be then intuitively understood as the sum of the expectation values of the observable $O$ evaluated in all possible quantum states of the $t\bar{t}$ pair in the region $\Pi$, with $p(M_{t\bar{t}},\hat{k})$ the probability of a given production process, proportional to the corresponding differential cross section. The description of the quantum state of the $t\bar{t}$ pair in terms of a density matrix instead of a wave function arises quite naturally in colliders: since internal degrees of freedom of the initial state (such as spin or color) cannot be controlled, the quantum state of the produced $t\bar{t}$ pair is described by an incoherent mixture resulting from the average over all possible initial states. 

In the same fashion of Eq. (\ref{eq:Expectationvalue}), we can define the total quantum state of the $t\bar{t}$ pair in $\Pi$ as
\begin{equation}\label{eq:TotalQuantumState}
\rho_\Pi\equiv\int_\Pi\mathrm{d}M_{t\bar{t}}\mathrm{d}\Omega~p(M_{t\bar{t}},\hat{k})\rho(M_{t\bar{t}},\hat{k}).
\end{equation}
As a two-qubit quantum state, $\rho_\Pi$ is determined by its coefficients $B^{\pm}_i,C_{ij}$. The motivation for considering $\rho_\Pi$ is that, as explained in Sec.~\ref{sec:QuantumTomography}, its spin polarizations and spin correlations can be extracted from measurements of accessible observables and hence, its quantum tomography can be implemented.

For the theoretical computation of $\rho_\Pi$, we use QCD perturbation theory at leading-order (LO). Higher-order corrections are expected to be small~\cite{Bernreuther2004}, while the physical picture at LO is simpler and neater, and the final results and conclusions still hold in the general case, as shown in Sec.~\ref{sec:experimental}.

At LO, only two initial states can produce a $t\bar{t}$ pair, which are precisely those described by Eq. (\ref{eq:partonreacitons}): a $q\bar{q}$ pair or a $gg$ pair. For fixed energy and top direction in the CM frame, each initial state $I=q\bar{q},gg$ gives rise to a different quantum state for the $t\bar{t}$ pair, characterized by a production spin density matrix $R^{I}(M_{t\bar{t}},\hat{k})$. The production spin density matrix $R$ for the total production process from $pp$ collisions is computed in terms of each partonic counterpart $R^{I}$ as
\begin{equation}\label{eq:Rtotal}
R(M_{t\bar{t}},\hat{k})=\sum_{I=q\bar{q},gg} L^{I}(M_{t\bar{t}})R^{I}(M_{t\bar{t}},\hat{k})
\end{equation}
with $L^{I}(M_{t\bar{t}})$ the so-called luminosity function, which  accounts for the incidence of each initial partonic state $I=q\bar{q},gg$ in the total process (see Ref.~\cite{Bernreuther1998} for the precise definition of the luminosity function in terms of PDF). They are numerically computed by using the NNPDF30LO PDF set~\cite{Ball:2014uwa}. In order to make sure the PDF set choice has a negligible impact on our calculations, we have examined other PDF sets as well. In particular, we have also used the CT10~\cite{Lai:2010vv} and the MSTW 2008~\cite{Martin:2009iq} PDF sets, obtaining similar results.

By using the expression of $\rho^{I}$ in terms of $R^{I}$, $\rho^{I}=R^{I}/4\tilde{A}^I$, we arrive at
\begin{equation}\label{eq:partonicstates}
\rho(M_{t\bar{t}},\hat{k})=\sum_{I=q\bar{q},gg} w_I(M_{t\bar{t}},\hat{k})\rho^{I}(M_{t\bar{t}},\hat{k}),
\end{equation}
finding that the probabilities $w_I$ are computed from the luminosities as
\begin{equation}\label{eq:partonicweights}
w_I(M_{t\bar{t}},\hat{k})=\frac{L^{I}(M_{t\bar{t}})\tilde{A}^I(M_{t\bar{t}},\hat{k})}{\sum_{J}L^{J}(M_{t\bar{t}})\tilde{A}^J(M_{t\bar{t}},\hat{k})} .
\end{equation}

\begin{figure}[tb!]\includegraphics[width=\columnwidth]{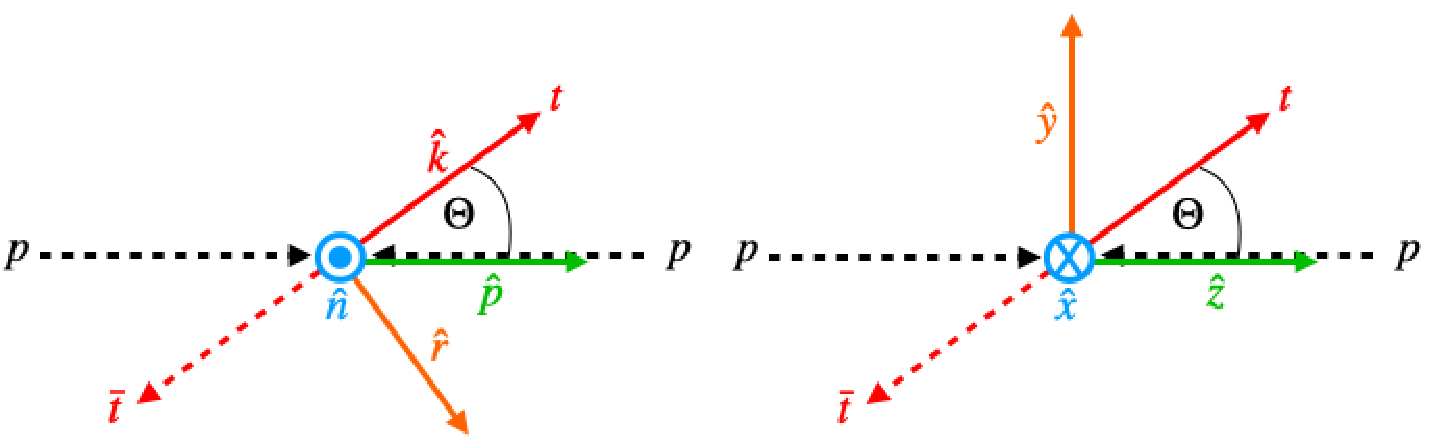}
\centering
     \caption{Different orthonormal basis in the CM frame considered in this work. Left: helicity basis. Right: beam basis.}
     \label{fig:Basis}
\end{figure}

The spin polarizations and correlations characterizing the production spin density matrices $R^{I}(M_{t\bar{t}},\hat{k})$ are computed in an orthonormal basis in the CM frame, the so-called helicity basis~\cite{Baumgart2013} $\{\hat{k},\hat{n},\hat{r}\}$, with $\hat{r}=(\hat{p}-\cos\Theta\hat{k})/\sin\Theta$ and $\hat{n}=\hat{r}\times\hat{k}$, $\hat{p}$ being the unitary vector in the direction of the proton beam and $\Theta$ the production angle with respect to the beam line, $\cos\Theta=\hat{k}\cdot \hat{p}$. A schematic representation of this basis is provided in left Fig.~\ref{fig:Basis}.

The production spin density matrix and all its coefficients are only functions of $\beta$ and $\cos\Theta$. Specifically, in the SM, the correlation matrix $\tilde{C}_{ij}$ is symmetric and $\tilde{B}_i^{+}=\tilde{B}_i^{-}$. Furthermore, at LO, the net polarizations vanish, $\tilde{B}_i^{\pm}=0$, and the spin in the $n$-axis is uncorrelated to the spin in the remaining directions, $\tilde{C}_{nr}=\tilde{C}_{nk}=0$. Thus, at LO, only $5$ parameters are needed to characterize the production spin density matrix: $\tilde{A},\tilde{C}_{kk},\tilde{C}_{nn},\tilde{C}_{rr},\tilde{C}_{kr}$. The values of these coefficients can be obtained analytically for each $R^I$ and are listed in Appendix~\ref{app:LODensityMatrix}, where the procedure to compute the production spin density matrix $R$ and the associated density matrices is also summarized.

\section{Entanglement in $t\bar{t}$ production} \label{sec:entanglement}

\begin{figure*}[tb!]
\begin{tabular}{@{}cc@{}}
    \includegraphics[width=0.49\textwidth]{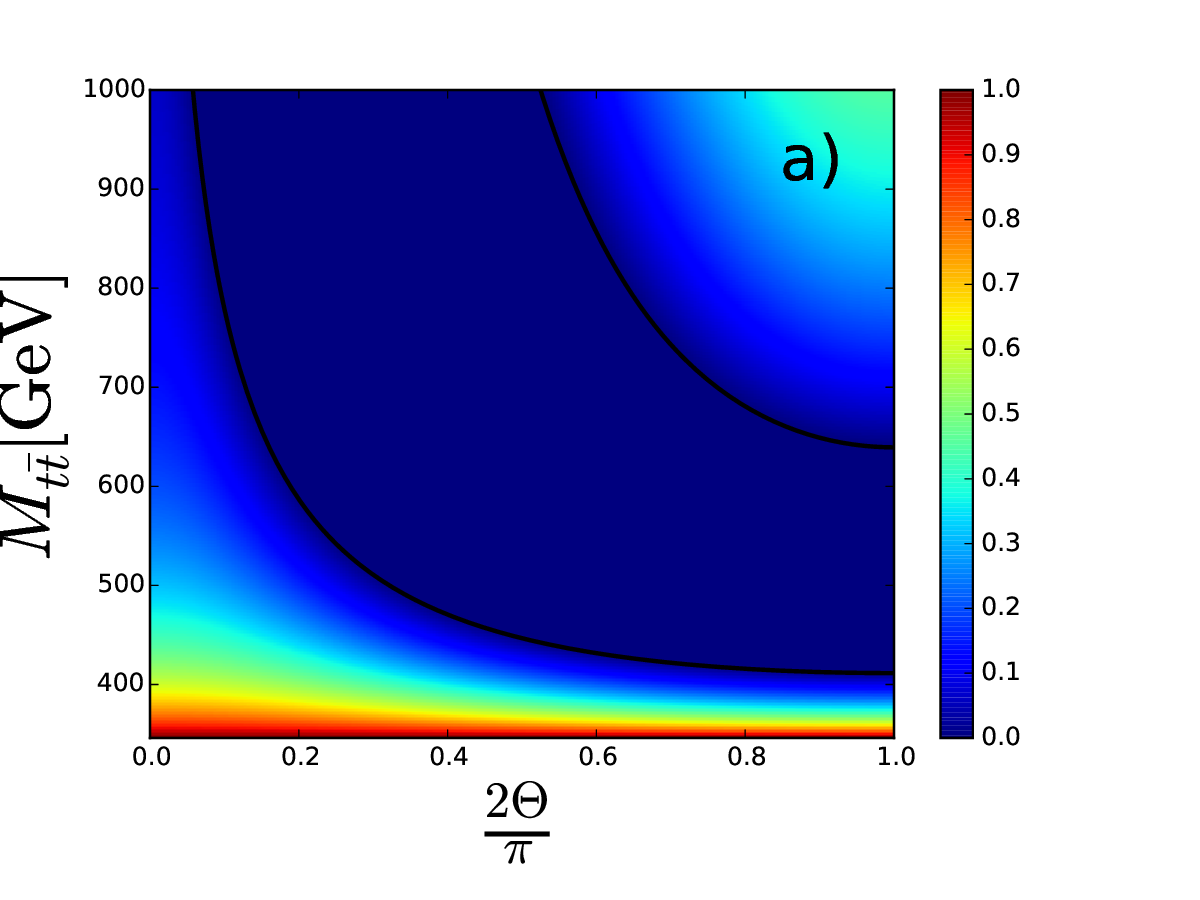} &
    \includegraphics[width=0.49\textwidth]{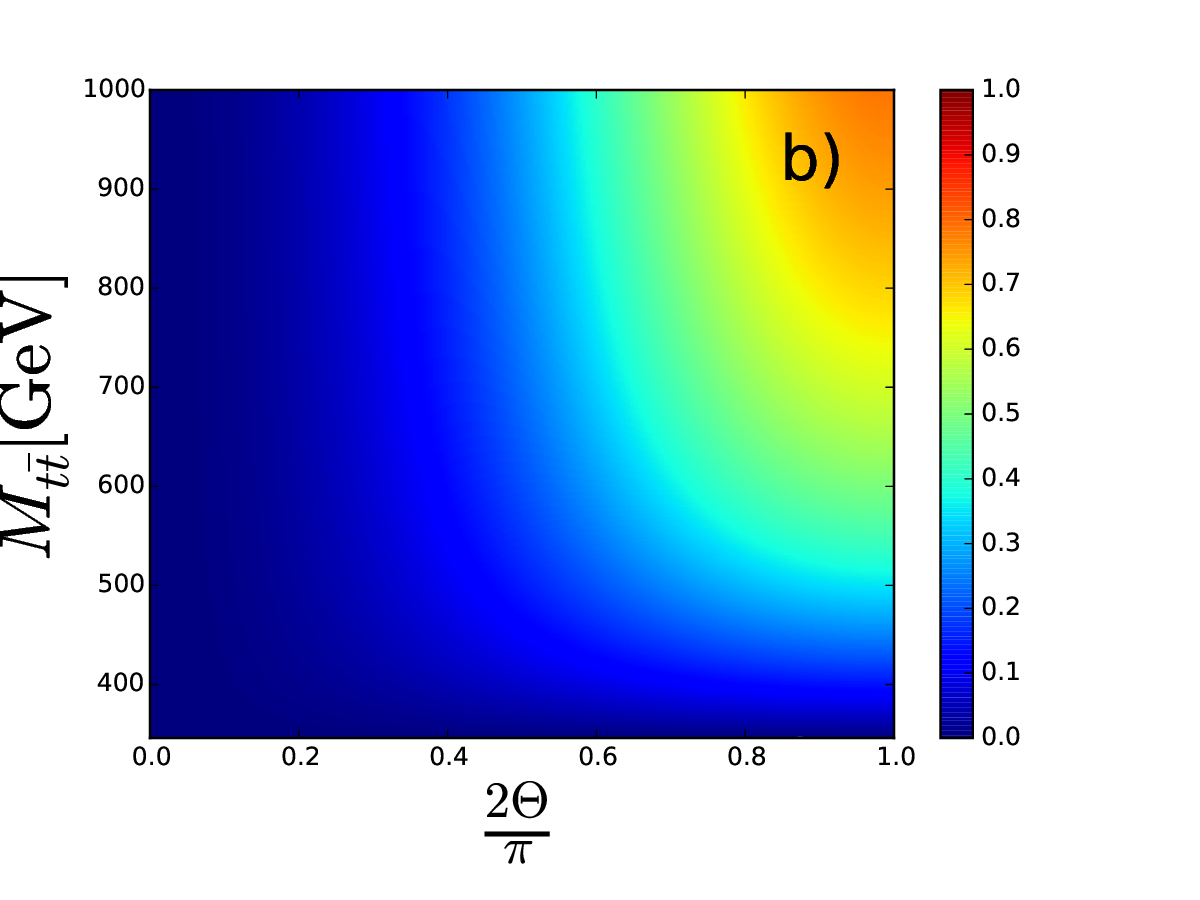} \\
    \includegraphics[width=0.49\textwidth]{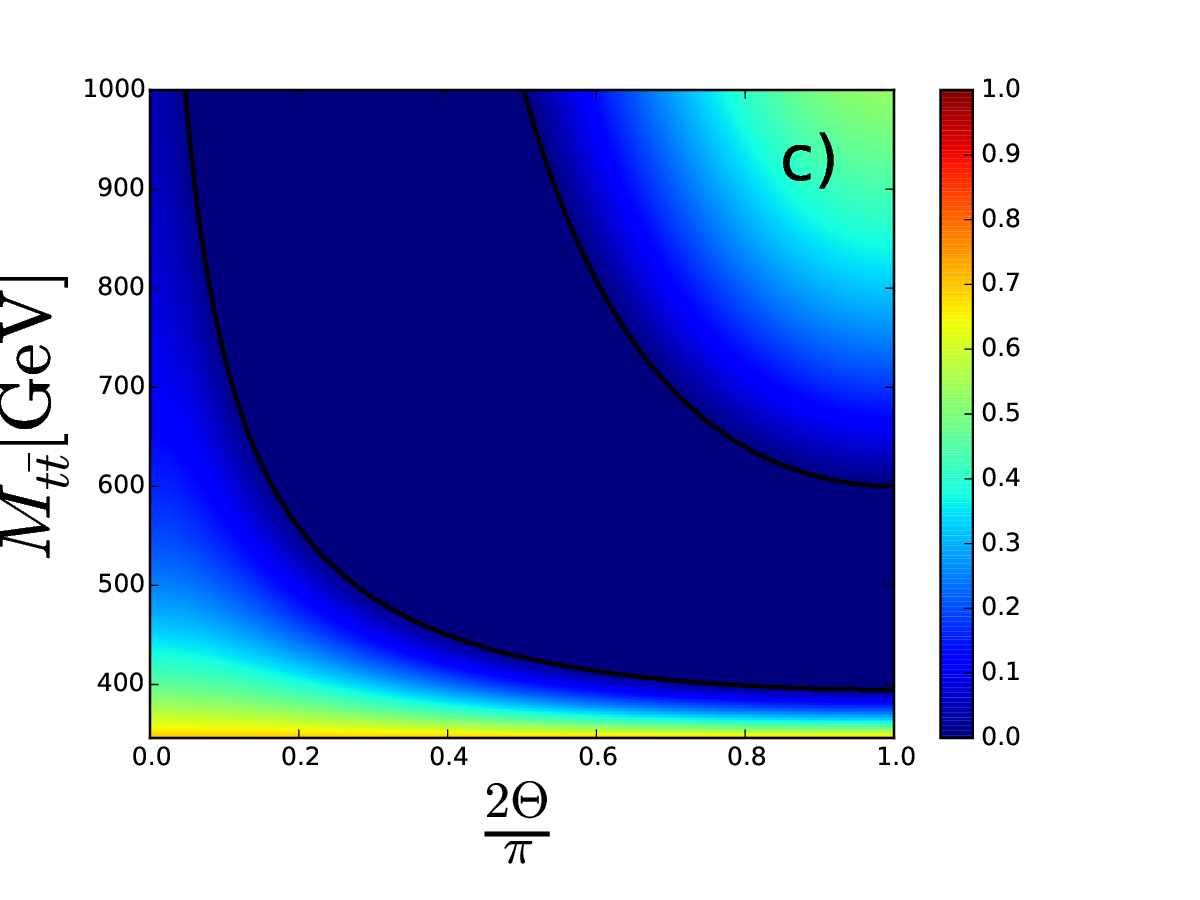} &
    \includegraphics[width=0.49\textwidth]{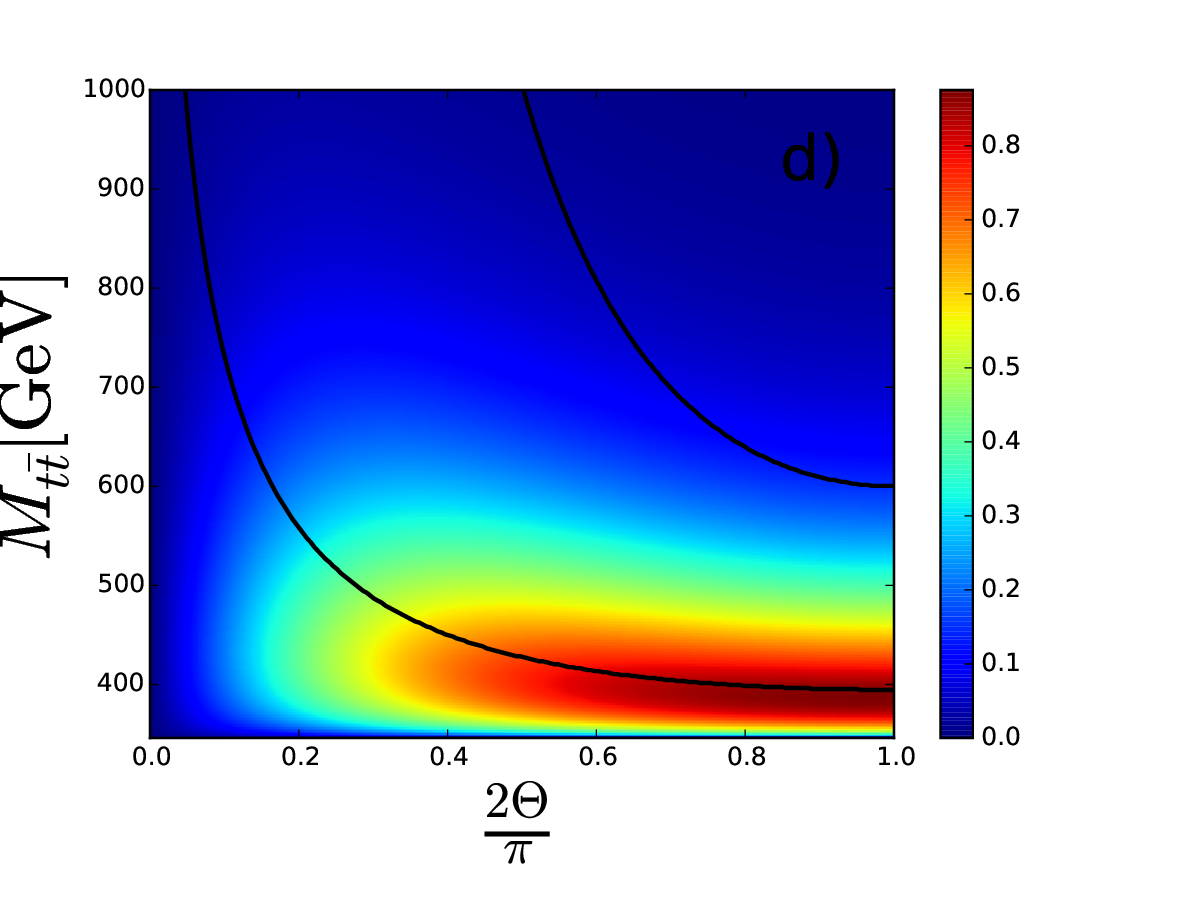}
\end{tabular}
\caption{Entanglement as a function of the invariant mass $M_{t\bar{t}}$ and the production angle $\Theta$ in the $t\bar{t}$ CM frame. All plots are symmetric under the transformation $\Theta\rightarrow \pi-\Theta$. Upper row: Concurrence of the spin density matrix $\rho^I(M_{t\bar{t}},\hat{k})$ of the $t\bar{t}$ pair resulting from the initial state $I=q\bar{q},gg$. a) $gg\rightarrow t\bar{t}$. Black lines represent the boundaries between separability and entanglement. b) $q\bar{q}\rightarrow t\bar{t}$. Lower row:  $t\bar{t}$ production at the LHC for $pp$ collisions at $\sqrt{s}=13$ TeV. Black lines represent the boundaries between separability and entanglement. c) Concurrence of the spin density matrix $\rho(M_{t\bar{t}},\hat{k})$. d) Differential cross section $\frac{d\sigma}{dM_{t\bar{t}}d\Theta}=2\pi\sin \Theta \frac{d\sigma}{dM_{t\bar{t}}d\Omega}$ in units of $\textrm{pb}/\textrm{GeV}~\textrm{rad}$.}
\label{fig:Concurrence}
\end{figure*}

Based on the definition of separability, if all sub-states $\rho(M_{t\bar{t}},\hat{k})$ are separable, the total quantum state of Eq. (\ref{eq:TotalQuantumState}) is also separable. The first natural step is then searching entanglement in the density matrices $\rho(M_{t\bar{t}},\hat{k})$, describing the quantum state of the $t\bar{t}$ pair for fixed energy and top direction in the CM frame. Since $\rho(M_{t\bar{t}},\hat{k})$ describes the spin quantum state of a pair of particles with well-defined momentum, its spin entanglement is a Lorentz invariant quantity~\cite{Gingrich2002,Peres2004}.

At LO, the density matrices $\rho(M_{t\bar{t}},\hat{k})$ are unpolarized and their correlation matrix is symmetric. By invoking the Peres-Horodecki criterion and applying the results of Appendix~\ref{app:criteria}, we find that
\begin{equation}\label{eq:Deltatt}
\Delta\equiv-C_{nn}+|C_{kk}+C_{rr}|-1>0
\end{equation}
is a necessary and sufficient condition for the presence of entanglement, with the concurrence simply given by
\begin{equation}\label{eq:ConcurrenceHelicity}
    C[\rho]=\frac{\max(\Delta,0)}{2} .
\end{equation}
Previous approaches for signaling entanglement in high-energy physics based on the entanglement entropy~\cite{Kharzeev2017,Martens2018,Tu2020} are not useful here since they are only valid for pure states. In terms of the partonic sub-states $\rho^{I}(M_{t\bar{t}},\hat{k})$, similar quantities $\Delta^{I}$ are defined. Specifically, for gluon fusion, we obtain
\begin{equation}
    \Delta^{gg}=\frac{2-4\beta^2(1+\sin^2\Theta)+2\beta^4(1+\sin^4\Theta)}{1+2\beta^2\sin^2\Theta-\beta^4(1+\sin^4\Theta)}
\end{equation}
if $\beta^2(1+\sin^2\Theta)<1$, while
\begin{equation}
    \Delta^{gg}=\frac{2\beta^4(1+\sin^4\Theta)-2}{1+2\beta^2\sin^2\Theta-\beta^4(1+\sin^4\Theta)}
\end{equation}
if $\beta^2(1+\sin^2\Theta)\geq 1$. Thus, $\rho^{gg}(M_{t\bar{t}},\hat{k})$ is separable in a finite region of phase space, with lower and upper  critical boundaries $\beta_{c1}(\Theta),\beta_{c2}(\Theta)$ between entanglement and separability
\begin{eqnarray}\label{eq:BetaCriticals}
\beta_{c1}(\Theta)&=&\sqrt{\frac{1+\sin^2\Theta-\sqrt{2}\sin\Theta}{1+\sin^4\Theta}}~,\\
\nonumber \beta_{c2}(\Theta)&=&\frac{1}{(1+\sin^4\Theta)^{\frac{1}{4}}}~.
\end{eqnarray}

The plot of the concurrence for $\rho^{gg}(M_{t\bar{t}},\hat{k})$ is shown in Fig.~\ref{fig:Concurrence}a. We can understand the presence of entanglement in the lower and upper regions of the plot from the nature of the $t\bar{t}$ production through gluon fusion. The spin polarizations of the gluon pair are allowed to align in different directions; at threshold (lower region of Fig. \ref{fig:Concurrence}a), this feature produces a $t\bar{t}$ pair in a spin-singlet state,
\begin{equation}\label{eq:singletgg}
\rho^{gg}(2m_t,\hat{k})=\ket{\Psi_0}\bra{\Psi_0},~\ket{\Psi_0}=\frac{\ket{\uparrow_{\hat{n}}\downarrow_{\hat{n}}}-\ket{\downarrow_{\hat{n}}\uparrow_{\hat{n}}}}{\sqrt{2}}
\end{equation}
with $\ket{\uparrow_{\hat{n}}},\ket{\downarrow_{\hat{n}}}$ the spin eigenstates along the direction $\hat{n}$. A spin-singlet state is maximally entangled, which explains the strong entanglement signature observed close to threshold. In the opposite limit of very high energies and production angles (upper right corner of Fig. \ref{fig:Concurrence}a), the produced $t\bar{t}$ pair is in a spin-triplet pure state,
\begin{equation}\label{eq:triplet}
\rho^{gg}(\infty,\hat{n}\times\hat{p})=\ket{\Psi_{\infty}}\bra{\Psi_{\infty}},~\ket{\Psi_{\infty}}=\frac{\ket{\uparrow_{\hat{n}}\downarrow_{\hat{n}}}+\ket{\downarrow_{\hat{n}}\uparrow_{\hat{n}}}}{\sqrt{2}}
\end{equation}
also maximally entangled.

On the other hand, for a $q\bar{q}$ initial state, the state is entangled in all phase space since
\begin{eqnarray}
\Delta^{q\bar{q}}=\frac{\beta^2\sin^2\Theta}{2-\beta^2\sin^2\Theta}\geq 0
\end{eqnarray}
This inequality is only saturated at $\Theta=0$ or at threshold, where $q\bar{q}$ reaction produces a $t\bar{t}$ pair with spins aligned along the beam axis in a correlated but separable mixed state,
\begin{equation}\label{eq:qqseparable}
\rho^{q\bar{q}}(2m_t,\hat{k})=\rho^{q\bar{q}}(M_{t\bar{t}},\hat{p})=\frac{\ket{\uparrow_{\hat{p}}\uparrow_{\hat{p}}}\bra{\uparrow_{\hat{p}}\uparrow_{\hat{p}}}+\ket{\downarrow_{\hat{p}}\downarrow_{\hat{p}}}\bra{\downarrow_{\hat{p}}\downarrow_{\hat{p}}}}{2}
\end{equation}
Thus, in these limits, the degree of entanglement is expected to be small, as can be seen in Fig. \ref{fig:Concurrence}b where we represent the concurrence of $\rho^{q\bar{q}}(M_{t\bar{t}},\hat{k})$.

We also see that in the opposite limit of very high energies and production angles, we reach again a maximally entangled state, as in Fig. \ref{fig:Concurrence}a. Indeed, in this limit the quantum state of the $t\bar{t}$ pair converges to the same state for $gg$ production, Eq. (\ref{eq:triplet}). The reason behind the convergence is the dominance of the orbital angular momentum contribution over the spin contribution.

With the help of the partonic processes, we compute the spin density matrix $\rho(M_{t\bar{t}},\hat{k})$ characterizing $t\bar{t}$ production at the LHC by Eq. (\ref{eq:partonicstates}). For the computation of the probabilities $w_I(M_{t\bar{t}},\hat{k})$ for each process, the CM frame of the $pp$ collisions is set to be $\sqrt{s}=13$~TeV, which corresponds to the latest data recorded~\cite{Aaboud:2019hwz,Sirunyan:2019lnl}. We analyze the presence of entanglement in $\rho(M_{t\bar{t}},\hat{k})$ in Fig. \ref{fig:Concurrence}c, while in Fig. \ref{fig:Concurrence}d the associated differential cross section is displayed. We note the strong resemblance between Figs.~\ref{fig:Concurrence}a,c, resulting from the dominance of gluon fusion in $t\bar{t}$ production at the LHC, i.e., $w_{q\bar{q}}(M_{t\bar{t}},\hat{k})\ll w_{gg}(M_{t\bar{t}},\hat{k})$~\cite{Bernreuther2004}. From Fig. \ref{fig:Concurrence}d, we also see that a strong signal is expected in the lower sub-region close to threshold where $\rho(M_{t\bar{t}},\hat{k})$ is entangled.

\section{Total quantum state}\label{sec:totalquantumstate}

\begin{figure*}[tb!]
\begin{tabular}{@{}cc@{}}
    \includegraphics[width=0.49\textwidth]{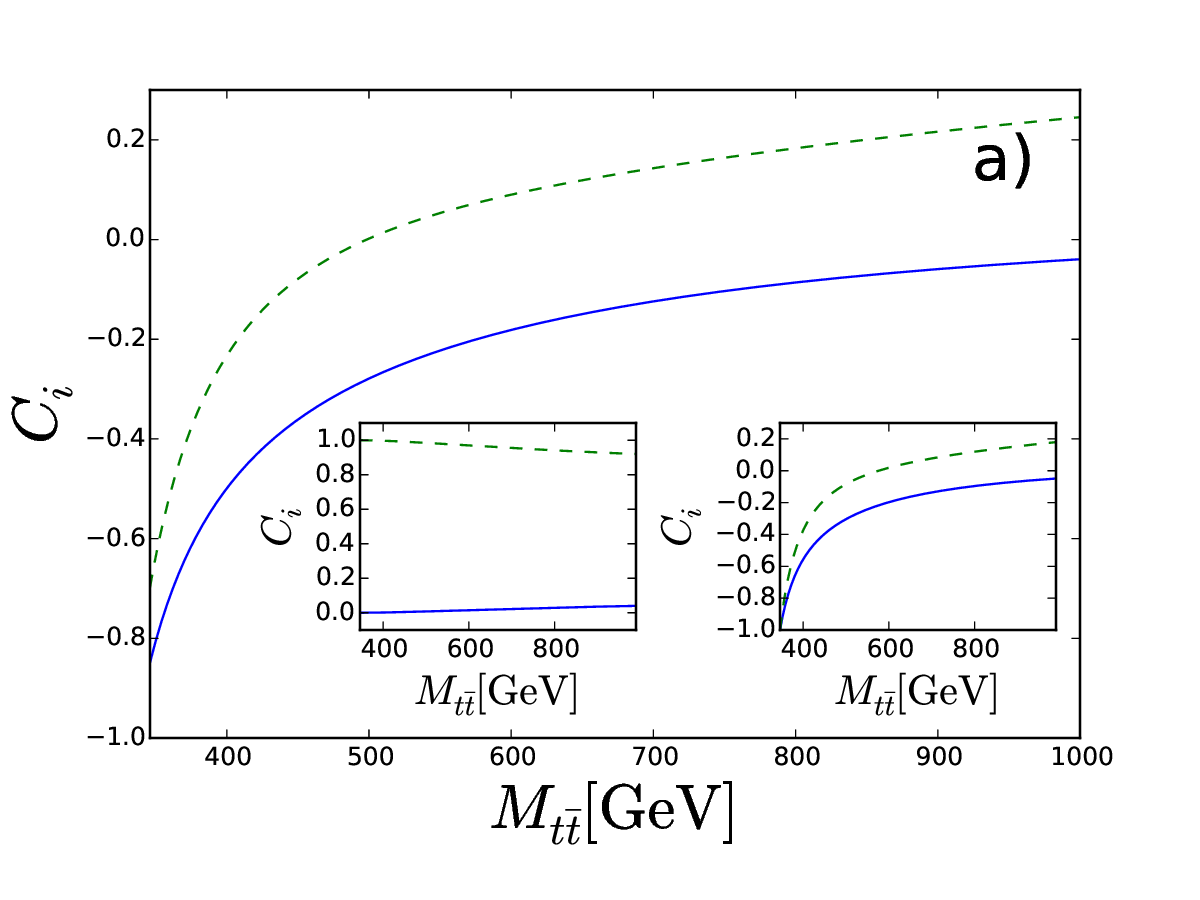} &
    \includegraphics[width=0.49\textwidth]{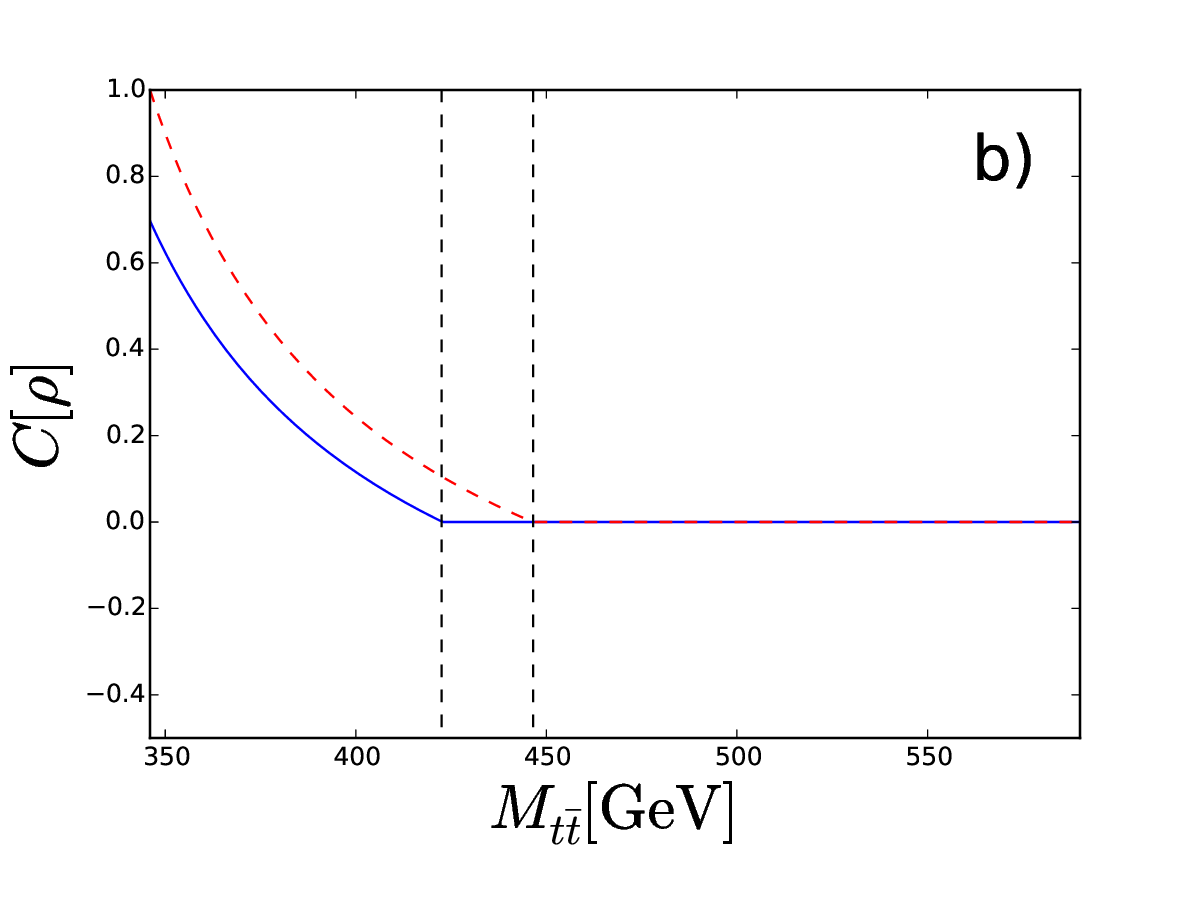} \\
    \includegraphics[width=0.49\textwidth]{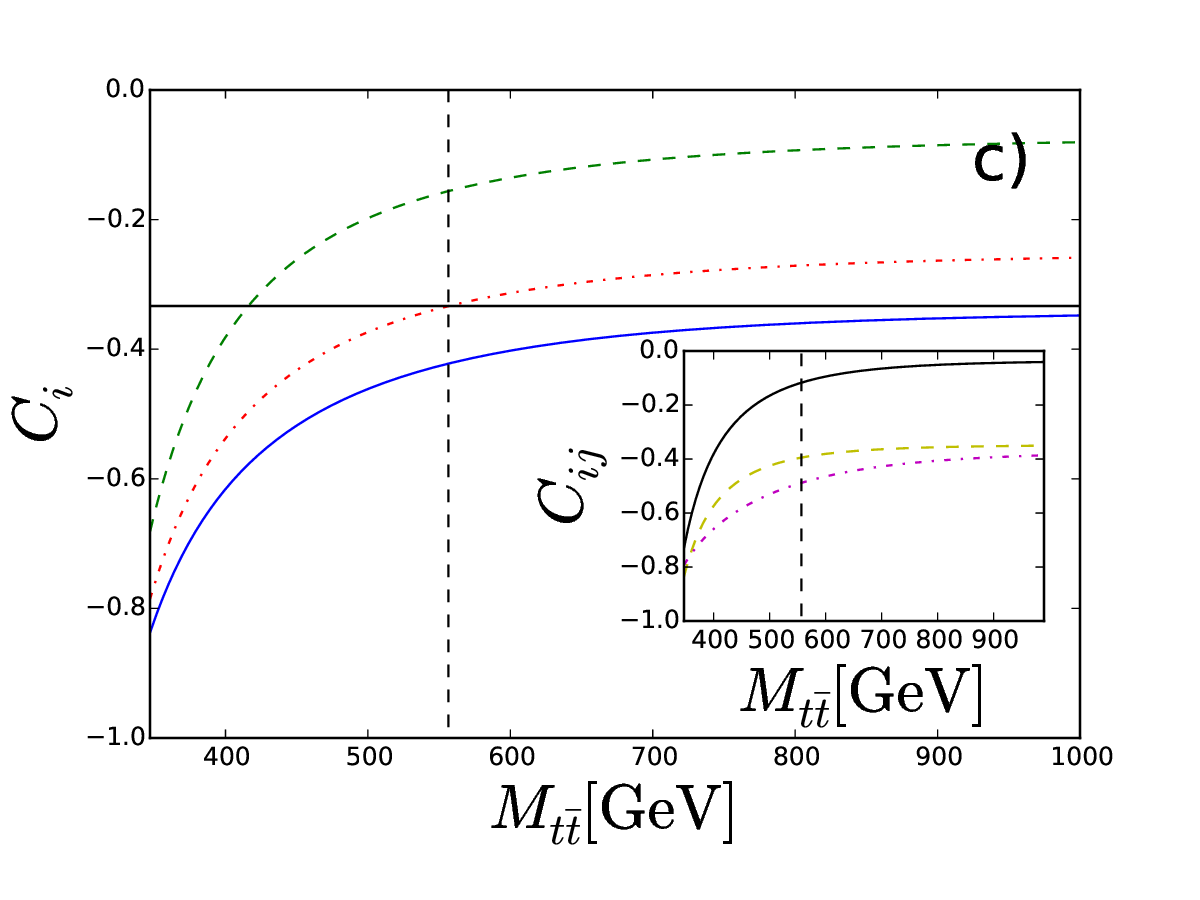} &
    \includegraphics[width=0.49\textwidth]{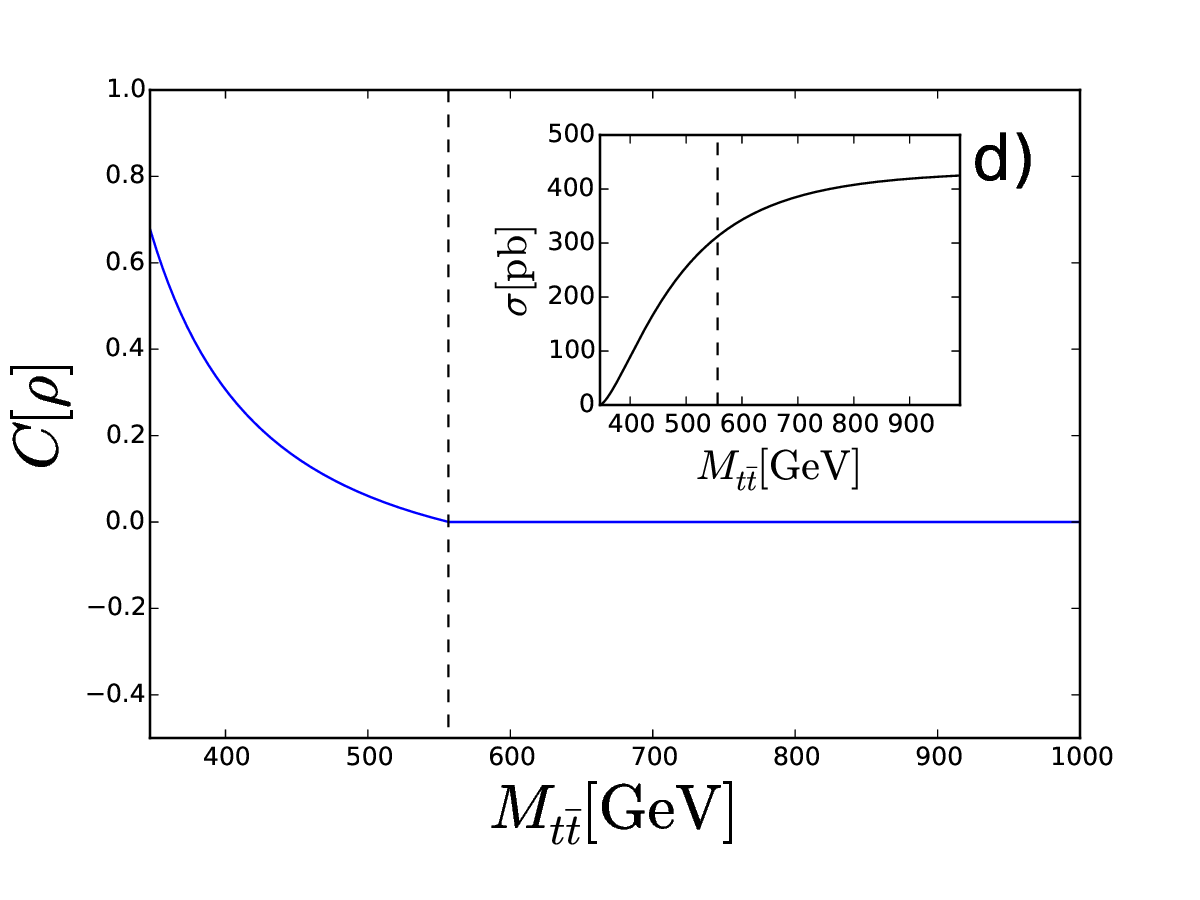}
\end{tabular}
\caption{Analysis of $t\bar{t}$ production at the LHC for $pp$ collisions at $\sqrt{s}=13$ TeV as a function of the invariant mass $M_{t\bar{t}}$. a) Spin correlations $C_{\perp}$ (solid blue) and $C_z$ (dashed green) for $\rho_{\Omega}$. Left and right insets: same but for $\rho_{\Omega}^{q\bar{q}},\rho_{\Omega}^{gg}$, respectively. b) Concurrence of $\rho_{\Omega}$ (solid blue) and $\rho_{\Omega}^{gg}$ (red dashed line). Vertical dashed lines signal the transition between entanglement and separability. c) Spin correlations integrated in the invariant mass window $[2m_t, M_{t\bar{t}}]$. Main plot: $C_{\perp}$ (solid blue), $C_{z}$ (dashed green), and $D$ (dashed-dotted red). The horizontal solid line and the vertical dashed line signal the entanglement limit $D=-1/3$. Inset: $C_{rr}$ (solid black), $C_{nn}$ (dashed yellow), and $C_{kk}$ (dashed-dotted purple). d) Main plot: Concurrence of the total quantum state $\rho(M_{t\bar{t}})$. Inset: integrated cross section in the window $[2m_t, M_{t\bar{t}}]$.}
\label{fig:CorrelationsFrame}
\end{figure*}

The results of the previous section suggest searching entanglement close to threshold. Specifically, we compute the total quantum state of Eq. (\ref{eq:TotalQuantumState}) by introducing a requirement in the mass spectrum, a common tool in the literature~\cite{Uwer2005,Bernreuther2015},
\begin{eqnarray}\label{eq:TotalQuantumStateMass}
\nonumber \rho(M_{t\bar{t}})&\equiv& \int^{M_{t\bar{t}}}_{2m_t}\mathrm{d}M\int\mathrm{d}\Omega~p(M,\hat{k})\rho(M,\hat{k})\\
&=&\int^{M_{t\bar{t}}}_{2m_t}\mathrm{d}M~p(M)\rho_{\Omega}(M)
\end{eqnarray}
where the integral limits mean that we only select events with invariant mass in the window $[2m_t,M_{t\bar{t}}]$, and $p(M_{t\bar{t}})$, $\rho_{\Omega}(M_{t\bar{t}})$ arise from the angular integration.

\subsection{Angular integration}\label{subsec:angularint}

Since it is obtained by averaging over all possible top directions, $\rho_{\Omega}(M_{t\bar{t}})$ needs to be computed in a \textit{fixed} spatial basis, making the \textit{co-moving} helicity basis useless for this purpose as it changes its orientation with the top direction for each individual production process. Specifically, due to the symmetry around the beam axis, we choose the so-called beam basis~\cite{Bernreuther2004} $\{\hat{x},\hat{y},\hat{z}\}$, with $\hat{z}=\hat{p}$ along the beam and $\hat{x},\hat{y}$ pointing transverse directions. A schematic representation of the beam basis is provided in right Fig. \ref{fig:Basis}.

The invariance under rotations around the beam axis implies that in this basis the spin correlation matrix is diagonal and satisfies $C_{ij}=\delta_{ij}C_j$, with $C_{x}=C_{y}\equiv C_{\perp}$. Thus, we only need $2$ parameters, the transverse and longitudinal spin correlations $C_{\perp},C_{z}$, in order to compute $\rho_{\Omega}(M_{t\bar{t}})$.

The spin density matrix $\rho_{\Omega}(M_{t\bar{t}})$ is computed in terms of its partonic counterparts $\rho_{\Omega}^{I}(M_{t\bar{t}})$, which are analytically calculated in Appendix~\ref{app:AngularAveraging}. They are related to each other in the same fashion of Eq. (\ref{eq:partonicstates}),
\begin{equation}\label{eq:partonicstatesangular}
\rho_{\Omega}(M_{t\bar{t}})=\sum_{I=q\bar{q},gg} w_I(M_{t\bar{t}})\rho_{\Omega}^{I}(M_{t\bar{t}})
\end{equation}
where the probabilities $w_I(M_{t\bar{t}})$ are given now by replacing $\tilde{A}^{I}(M_{t\bar{t}},\hat{k})$ by its angular-averaged counterpart $\tilde{A}^{I}(M_{t\bar{t}})$ in Eq. (\ref{eq:partonicweights}); the same relation applies between the probabilities $p(M_{t\bar{t}})$ and $p(M_{t\bar{t}},\hat{k})$. The transverse and longitudinal spin correlations are represented in Fig. \ref{fig:CorrelationsFrame}a.

With respect to entanglement, due to the symmetry around the $z$-axis of the states $\rho_{\Omega}$, the Peres-Horodecki criterion is equivalent to $\delta>0$ [see Eq. (\ref{eq:PeresHorodeckiAxialLO}) and ensuing discussion], with
\begin{equation} \label{eq:PeresHorodecki}
\delta\equiv-C_{z}+|2C_{\perp}|-1
\end{equation}
and the concurrence being given by $C[\rho_{\Omega}]=\max(\delta,0)/2$.

In terms of the partonic sub-states $\rho_{\Omega}^{I}(M_{t\bar{t}})$, we have
\begin{equation}
\delta^{q\bar{q}}(M_{t\bar{t}})=\frac{-2+2\frac{\beta^2}{3}+\frac{4}{15}\left(1-\sqrt{1-\beta^2}\right)^2}{1-\frac{\beta^2}{3}}<0
\end{equation}
so $\rho_{\Omega}^{q\bar{q}}$ is completely separable. In contrast, for gluon fusion we have that in the energy range of interest $C^{gg}_{\perp}(M_{t\bar{t}})<0$ (the sign crossover is only produced at very high energies $\beta=\beta_{\delta}\simeq0.970$), and then
\begin{eqnarray}
\delta^{gg}(M_{t\bar{t}})&\simeq&-C_{z}-2C_{\perp}-1=-\textrm{tr}[\mathbf{C}^{gg}]-1\\
&=&
\nonumber \frac{150-62\beta^2-\left(136-76\beta^2+4\beta^4\right)\frac{\textrm{atanh}(\beta)}{\beta}}{-59+31\beta^2+(66-36\beta^2+2\beta^4)\frac{\textrm{atanh}(\beta)}{\beta}}
\end{eqnarray}
From this expression, we compute the critical top velocity $\beta_c\simeq 0.632$ below which the state $\rho^{gg}_{\Omega}$ is still entangled, with the associated critical mass being $M_{c}=2m_t/\sqrt{1-\beta^2_c}\simeq2.58m_t\simeq446~\mathrm{GeV}$.

The entanglement loss for both quark and gluon processes arises due to the statistical average over all possible top directions. However, close to threshold, gluon fusion produces a $t\bar{t}$ pair in a spin singlet, invariant under rotations, and thus unaffected by the angular average, keeping the entanglement.

We can also obtain these results from the spin correlations in the helicity basis, given by Eqs. (\ref{eq:AngularAveragedCorrelationsHelicityqq}), (\ref{eq:AngularAveragedCorrelationsHelicitygg}). Indeed, since the relation $C^{gg}_{kk}+C^{gg}_{rr}<0$ is satisfied in the energy range where the state is entangled, we have that $\delta^{gg}=\Delta^{gg}=-\textrm{tr}[\mathbf{C}^{gg}]-1$. In actuality, any orthonormal basis serves for characterizing entanglement due to the invariance of the trace of the correlation matrix, $\textrm{tr}[\mathbf{C}]=2C_{\perp}+C_z=C_{rr}+C_{nn}+C_{kk}=\braket{\mathbf{\sigma}\cdot \bar{\mathbf{\sigma}}}$, reflecting the rotational symmetry of the spin-singlet state.

The dominance of $gg$ processes at the LHC implies that $\rho_{\Omega}$ is still entangled in some energy range, in which $\delta=\Delta=-\textrm{tr}[\mathbf{C}]-1>0$. The concurrences of $\rho_{\Omega},\rho_{\Omega}^{gg}$ are represented in Fig. \ref{fig:CorrelationsFrame}b.

\subsection{Mass integration}\label{subsec:massint}

We finally compute here the total quantum state of the system $\rho(M_{t\bar{t}})$ by performing the mass integration in Eq. (\ref{eq:TotalQuantumStateMass}) with the help of the angular results of the previous subsection. Since we are only integrating in the mass window, the properties of $\rho(M_{t\bar{t}})$ are similar to those of $\rho_{\Omega}(M_{t\bar{t}})$. In particular, the total quantum state of the system is also computed in the beam basis, characterized by its transverse and longitudinal spin correlations $C_{\perp},C_{z}$, represented in Fig. \ref{fig:CorrelationsFrame}c. For comparison, the inset shows the average within the same region of phase space of the spin correlations in the more usual helicity basis.

In Fig. \ref{fig:CorrelationsFrame}d we analyze the entanglement of $\rho(M_{t\bar{t}})$, where we can see that, by imposing a requirement on the invariant mass, we get a strong signal (as shown by the integrated cross section in the inset) while clearly detecting entanglement. A more extensive experimental analysis is provided in Sec.~\ref{sec:experimental}.

We note that the critical mass here is significantly larger than that of the angular-averaged sub-states $\rho_{\Omega}(M_{t\bar{t}})$, Fig. \ref{fig:CorrelationsFrame}b. This increase arises from the fact that the total quantum state $\rho(M_{t\bar{t}})$ is a sum of the sub-states $\rho_{\Omega}(M_{t\bar{t}})$. Therefore, one needs to go to higher energies to include a sufficient amount of separable states to dilute the contribution of those entangled in order to make the total state $\rho(M_{t\bar{t}})$ separable. However, if the integration window in the mass spectrum is entirely placed in the region of separability of $\rho_{\Omega}(M_{t\bar{t}})$, no entanglement would be detected.

\section{Quantum Tomography}\label{sec:QuantumTomography}

In this section, we design an experimental protocol to implement the quantum tomography of the total quantum state of the $t\bar{t}$ pair, built on well-established techniques.

Due to the large width of the top quark, $\Gamma_t\sim 1~\textrm{GeV}$, the $t\bar{t}$ pair has a very short lifetime and quickly decays, well before any other processes such as hadronisation or spin decorrelation could play a role. The spin information of the $t\bar{t}$ pair is then immediately transferred to the decay products. Specifically, when both the top/antitop decay products contain an antilepton/lepton ($l^+l^-$) pair, the angular differential cross section characterizing the process is given by~\cite{Baumgart2013}:
\begin{equation}\label{eq:LeptonicCrossSection}
\frac{1}{\sigma}\frac{\mathrm{d}\sigma}{\mathrm{d}\Omega_{+}\mathrm{d}\Omega_{-}}=\frac{1+\mathbf{B}^{+}\cdot\hat{\mathbf{q}}_{+}-\mathbf{B}^{-}\cdot\hat{\mathbf{q}}_{-}
-\hat{\mathbf{q}}_{+}\cdot \mathbf{C} \cdot\hat{\mathbf{q}}_{-}}{(4\pi)^2}
\end{equation}
where $\hat{\mathbf{q}}_{\pm}$ are the lepton (antilepton) directions in each one of the parent top (antitop) rest frames, and $\Omega_{\pm}$ the corresponding solid angles. The vectors $\mathbf{B}^{\pm}$ and the matrix $\mathbf{C}$ are precisely the top/antitop spin polarizations and the spin correlation matrix, respectively.

From the kinematic reconstruction of each event, the $t\bar{t}$ and $l^+l^-$ pairs momenta are determined, so the spin polarizations and correlations can be obtained in any basis. Indeed, this kind of measurement has already been carried out by the CMS collaboration at the LHC~\cite{Sirunyan:2019lnl}, where $\mathbf{B}^{\pm},\mathbf{C}$ were obtained in the helicity basis, with no restrictions on $t\bar{t}$ phase space. However, as discussed in Sec.~\ref{sec:totalquantumstate}, the helicity basis is not valid for reconstructing the total quantum state of the system since the latter already represents an average over all possible top directions. As a result, the quantum tomography can only be implemented from the measurement of the spin polarizations and correlations in a fixed spatial basis.

We propose a protocol to implement the quantum tomography of the $t\bar{t}$ pair based on the experimental scheme described above. In particular, we propose to perform the measurement of the spin polarizations and correlations using the fixed beam basis instead of the more usual helicity basis, also imposing an upper cut $M_{t\bar{t}}$ in the invariant mass spectrum of the $t\bar{t}$ pair. In this way, the measured $\mathbf{B}^{\pm},\mathbf{C}$ are the proper coefficients characterizing the total quantum state $\rho(M_{t\bar{t}})$ of Eq. (\ref{eq:TotalQuantumStateMass}), so its reconstruction is performed. These measurements are to be complemented with a specific maximum-likehood estimation~\cite{Hradil1997} for the parameters $\mathbf{B}^{\pm},\mathbf{C}$ that ensure that the reconstructed quantum state is a nonnegative operator. In other words: quantum tomography is not just an ensemble of independent expectation values but rather a method to reconstruct the total quantum state, a physical object, which as a result imposes further restrictions on the values of the measurements.

At LO, and by assuming symmetry around the beam axis, the measurement of only $2$ parameters, the transverse and longitudinal spin correlations $C_{\perp}(M_{t\bar{t}}),C_{z}(M_{t\bar{t}})$, is needed to perform the quantum tomography of $\rho(M_{t\bar{t}})$.

In general, by only assuming symmetry around the beam axis, the quantum tomography of the $t\bar{t}$ pair requires the measurement of just $4$ parameters, $B^{\pm}_{z}, C_{\perp}, C_z$, with $B^{\pm}_{z}$ the spin polarizations along the beam axis. In fact, even without any assumption on the specific form of $\rho(M_{t\bar{t}})$, by measuring all the $15$ parameters $B^{\pm}_{i},C_{ij}$ in the beam basis, the full quantum tomography can always be performed. A summary of the parameters needed to be measured in order to perform the quantum tomography of the $t\bar{t}$ pair is presented in Table~\ref{table}. We note that the presented scheme is simpler and goes beyond the general approach discussed in Ref.~\cite{Martens2018}, which would not be able to fully reconstruct the total quantum state of the $t\bar{t}$ pair.

\begin{table}[h]
\begin{center}
\begin{tabular}[c]{|c|c|c|}
\hline
Assumption & Coefficients & \# parameters\\
\hline
Symmetry and LO & $C_{\perp},C_{z}$ & 2 \\
\hline
Symmetry & $B^{\pm}_{z},C_{\perp},C_{z}$ & 4 \\
\hline
None & $B^{\pm}_{i},C_{ij}$ & 15 \\
\hline
\end{tabular}
\caption{Summary of the parameters needed to be measured in order to perform the quantum tomography of the $t\bar{t}$ pair for different assumptions on the form of $\rho(M_{t\bar{t}})$. ``Symmetry'' denotes symmetry around the beam axis.}
\label{table}
\end{center}
\end{table}

\section{Experimental entanglement detection}\label{sec:experimental}

\begin{figure*}[tb!]
\begin{tabular}{@{}cc@{}}
    \includegraphics[width=0.49\textwidth]{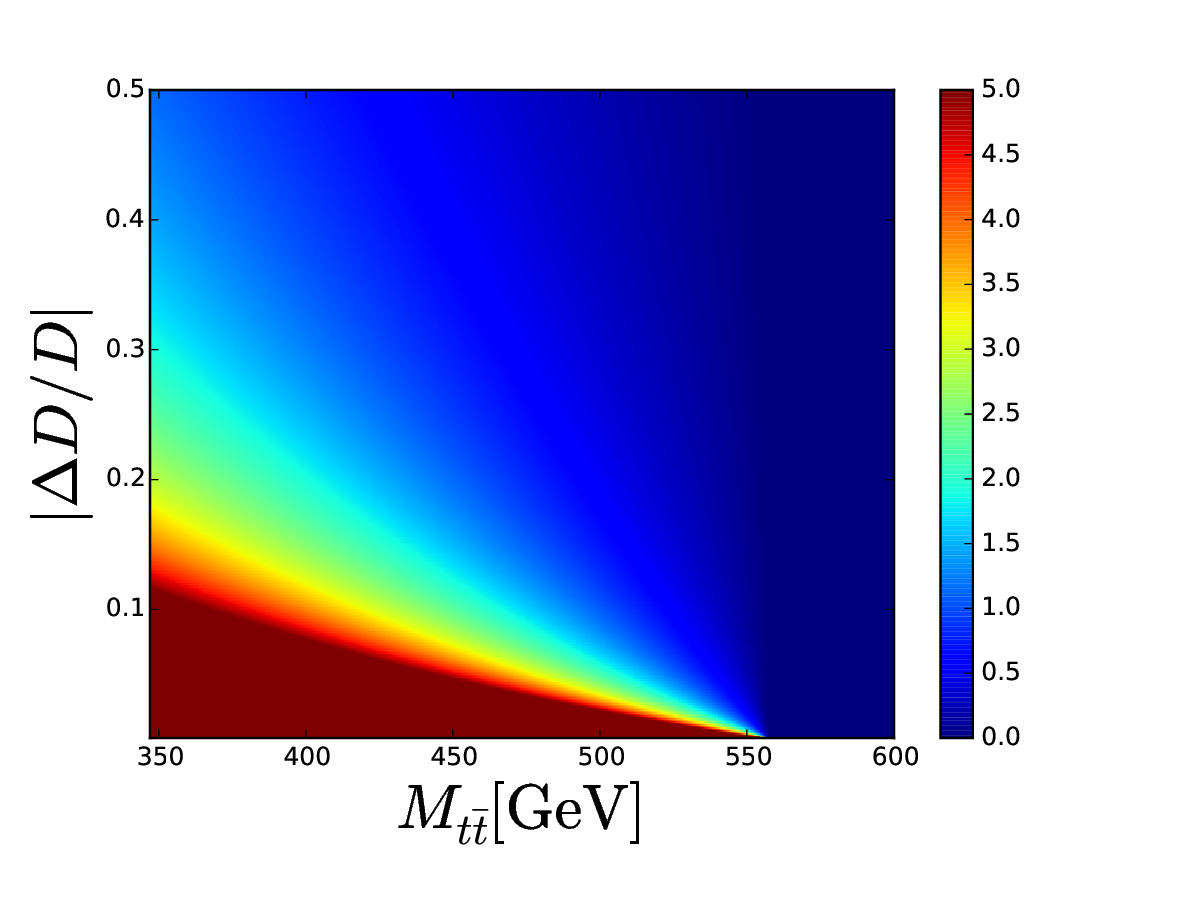} &
    \includegraphics[width=0.49\textwidth]{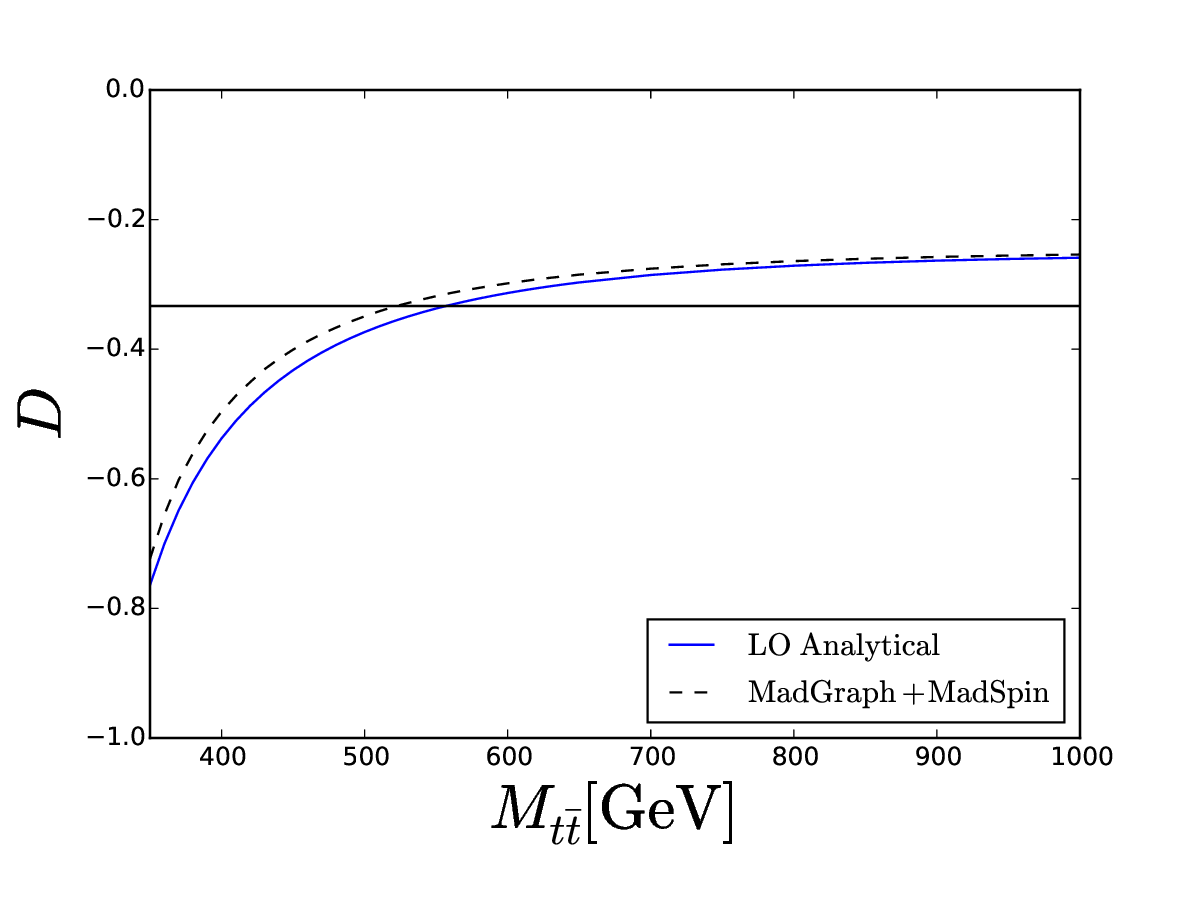}
\end{tabular}
\caption{Left: Statistical deviation from the null hypothesis ($D = -1/3$) for different assumptions of relative uncertainty on $D$. The contour shows the number of measurement uncertainties differing between the measured value of $D$ and the null hypothesis, $n_{\Delta}$.
Right: The value of $D$ within the mass window $[2m_t, M_{t\bar{t}}]$.
The LO analytical values are calculated using the methods presented in this work, while the {\sc MadGraph + MadSpin} values are calculated numerically by using Monte Carlo simulation. The horizontal line represents the critical value $D=-1/3$.}
\label{fig:Discovery}
\end{figure*}

For the experimental detection of entanglement, one can always implement the quantum tomography of the total quantum state following the protocol developed in the previous section, and study its separability (for instance, by computing the concurrence). However, interestingly, at the LHC a direct experimental entanglement signature is provided by the measurable observable
\begin{equation}\label{eq:ExperimentalPH}
    D=\frac{\textrm{tr}[\mathbf{C}]}{3}=-\frac{1+\delta}{3},
\end{equation}
which can be extracted from the differential cross section characterizing the angular separation between the leptons
\begin{equation}\label{eq:DMeasure}
\frac{1}{\sigma} \frac{d\sigma}{d\cos\varphi}=\frac{1}{2}(1 - D\cos\varphi)
\end{equation}
where $\varphi$ is the angle between the lepton directions in each one of the parent top and antitop rest frames [see also Eq. (\ref{eq:LeptonicCrossSection})]. This quantity, also represented in Fig. \ref{fig:CorrelationsFrame}c, provides a simple entanglement criterion since the condition $\delta>0$ translates into $D<-1/3$. The concurrence is also readily measured from $D$ as $C[\rho]=\max(-1-3D,0)/2$.

The detection of entanglement is more non-trivial than could naively be expected since, even though entanglement is present in a wide region of phase space, the statistical averaging over all possible directions induces the necessity of a selection in the mass spectrum. This observation was already evident from the recent measurement of the CMS collaboration~\cite{Sirunyan:2019lnl}, in which it was obtained $D=-0.237\pm0.011>-1/3$ without any requirements on the mass window.

Our proposal for the experimental detection of entanglement is similar to the quantum tomography protocol developed in the previous section. The idea is to measure $D$ from the cross section of Eq. (\ref{eq:DMeasure}), also applying an upper cut in the invariant mass spectrum. Left Fig. \ref{fig:Discovery} presents an experimental perspective for the entanglement detection at the LHC. The null hypothesis is defined to be the upper limit where $D$ does not signal entanglement, $D = -1/3$. We represent the number of measurement uncertainties $n_\Delta$ differing between the expected measurement and the null hypothesis,
\begin{equation}
    n_{\Delta}\equiv \max\left[\frac{D+1/3}{\Delta D},0\right],
\end{equation}
as a function of the upper cut in the invariant mass spectrum, $M_{t\bar{t}}$, and the relative uncertainty, $\lvert \Delta D / D \rvert$, where $D$ is the expected value [computed theoretically from Eq. (\ref{eq:ExperimentalPH})] and $\Delta D$ the uncertainty of the measurement.

Any measurement with $n_\Delta>5$ implies a detection of entanglement within $5$ statistical deviations ($5\sigma$). In particular, the recent measurement of $D$ above quoted has a relative uncertainty of $4.6\%$~\cite{Sirunyan:2019lnl}, allowing a measurement of entanglement with more than $5\sigma$. We note high enough statistics is expected even with hard selection on the $t \bar{t}$ invariant mass spectrum. For instance, with a total integrated luminosity of $139~\textrm{fb}^{-1}$, which is the current data recorded by the LHC, we deduce that a selection of $[2m_t,M_{t\bar{t}}]$ with $M_{t\bar{t}} = 450$ GeV leaves $\sim 5 \cdot 10^4$ events~\cite{Aaboud:2019hwz}, accounting for selection efficiency and detector acceptance. For this selection, entanglement can be measured within $5\sigma$ if the relative uncertainty is up to $6\%$. Lower selection of $M_{t\bar{t}}$ decrease the value of $D$, allowing higher relative uncertainties to have similar statistical significance.

We note that a full estimation of the background processes at the LHC is beyond the scope of this work. Nevertheless, the estimation above is done using a result of an analysis performed by the CMS collaboration~\cite{Sirunyan:2019lnl}, while similar analyses have been performed by the ATLAS and CMS collaborations~\cite{ATLAS:2012ao,Aad:2014pwa,Aad:2014mfk,Aad:2015bfa,Aaboud:2016bit,Aaboud:2019hwz,Chatrchyan:2013wua,Khachatryan:2015tzo,Khachatryan:2016xws}. Those analyses estimated the background processes for the suggested measurement and clearly showed that the measurable quantities are controllable at the LHC. 

Although our calculation is restricted to LO, in general, the criterion $D<-1/3$ still provides a sufficient condition for entanglement without any assumption on the specific form of $\rho(M_{t\bar{t}})$ since it is equivalent to $\textrm{tr}[\mathbf{C}]<-1$ [see Eq. (\ref{eq:DeltaTraceGeneralized})]. Hence, the range of values $-1\leq D<-1/3$ represents a genuine non-classical feature, which can be qualitatively understood from the fact that the classical average of the scalar product of two vectors with unit length is never larger than one, $|\textrm{tr}[\mathbf{C}]|=|\braket{\mathbf{\sigma}\cdot \bar{\mathbf{\sigma}}}|\leq 1$. Indeed, we can regard $D<-1/3$ as the violation of a Cauchy-Schwarz inequality, a well-known entanglement signature in other fields such as quantum optics, condensed matter or analog Hawking radiation~\cite{Walls2008,Wolk2014,deNova2014}. We note that a similar entanglement criterion was derived in the context of Heisenberg spin chains~\cite{Schliemann2003}.

In order to validate our results, we have also calculated $D$ numerically by using {\sc MadGraph}~\cite{Alwall:2011uj} and {\sc MadSpin}~\cite{Artoisenet:2012st,Frixione:2007zp}.
As noted earlier, we set $\sqrt{s}=13$~TeV, corresponding to the latest data recorded at the LHC. We find similar results to the analytical calculation, as shown in right Fig. \ref{fig:Discovery}. We note that spin correlation effects are calculated by {\sc MadSpin} at tree-level accuracy, but NLO corrections are small, see, e.g., Fig. 3 in Ref.~\cite{Artoisenet:2012st}.

\section{Conclusions and outlook}\label{sec:conclusions}

In the present study of entanglement between a pair of $t\bar{t}$ quarks, we show that it can be detected with high statistical significance, using the currently recorded data by Run 2 at the LHC. By analyzing the contribution of each initial state to the total $t\bar{t}$ production, we identify gluon fusion close to threshold as the key entanglement source since it gives rise to a $t\bar{t}$ pair in a spin-singlet state that, due to its invariance under rotations, keeps the entanglement after averaging over all possible directions. Remarkably, the invariant character of the spin-singlet state and the dominance of gluon fusion in $t\bar{t}$ production at the LHC also imply that the angular separation between the leptonic decay products provides a direct experimental entanglement signature. Thus, entanglement can be observed from the measurement of just one accessible magnitude. The resulting experiment would provide the entanglement detection at the highest-energy scale ever made.

We stress the non-trivial character of the entanglement detection since it can only be performed in a reduced region of phase space close to threshold, which requires the implementation of restrictions in the mass spectrum for the observation. Indeed, a recent measurement by the CMS collaboration~\cite{Sirunyan:2019lnl} was not able to show entanglement due to the absence of such restrictions.

Moreover, we provide a simple experimental scheme for the quantum tomography of the $t\bar{t}$ pair, based on well-established techniques~\cite{Sirunyan:2019lnl} that allow to measure the spin correlations and polarizations of the $t\bar{t}$ pair from the cross section of the leptons arising from their dileptonic decay. Specifically, we propose to perform the same type of measurement but with an upper cut in the mass spectrum, and using the fixed beam basis instead of the usual helicity basis for the spin characterization, since the reconstruction of the quantum state requires the use of a spin orthonormal basis that does not change its orientation in each individual event. Under the realistic physical assumption of symmetry around the beam axis, our quantum tomography protocol requires the measurement of solely $4$ parameters: the transverse and longitudinal spin correlations between the $t\bar{t}$ pair, and their longitudinal spin polarizations along the beam axis.

The quantum tomography provides full knowledge of the quantum state of the system, going beyond the detection of entanglement, and the design of an experimental protocol for its implementation represents the other main result of the work. For instance, one could perform the quantum tomography of the $t\bar{t}$ pair in the region of separability at high energies in order to study the so-called quantum discord~\cite{Ollivier2001}.  

Our calculations have been developed within an analytical LO approximation that provides a simple and neat picture of the underlying physics. A full quantitative state-of-the-art calculation, including higher-order corrections, along with a full estimation of the background processes for the experimental detection, is beyond the scope of this work and left for future analysis. Nevertheless, such an analysis would only introduce quantitative corrections to the numerical results, and would not modify the main conclusions of the work. In particular, the proposed experimental entanglement criterion $D<-1/3$ and the developed quantum tomography protocol are valid in general, and in particular to arbitrary order in perturbation theory.

From a global perspective, the work implements a number of canonical techniques of quantum information, such as the quantum tomography or the concurrence, in a high-energy context, opening the prospect of studying quantum information aspects at high-energy colliders. The genuine relativistic behavior, the exotic character of the interactions and symmetries involved, and the fundamental nature of this environment make it especially attractive for such purpose.

From the high-energy perspective, a natural and intriguing continuation of this work is the exploration of new physics beyond the standard model through its possible effects on the quantum state of the $t\bar{t}$ pair, measurable through quantum tomography, or even through a modification of the concept of entanglement itself. 

Another line of research is the extension of the present study to other existing high-energy colliders, such as the Tevatron. In this respect, a very interesting perspective for the future is provided by the Future Circular Collider (FCC)~\cite{Benedikt2020}, where $t\bar{t}$ pairs will be produced from the collision of positron-electron ($e^{+}e^{-}$) pairs. The possibility of controlling the spin degrees of freedom of the initial $e^{+}e^{-}$ state makes the FCC a very rich and promising system.

Finally, the techniques developed in this work can be exported to the study of entanglement in other high-energy processes involving qubits and/or qutrits (as, for instance, massive spin-$1$ particles like the electroweak bosons $W^{\pm},Z^0$). One thing is sure: the high-energy physics environment is an extremely interesting arena for the study of quantum information theory and other fundamental aspects of quantum mechanics. 

\acknowledgements

We thank J. J. Garc\'ia-Ripoll, F. Sols, Y. Rozen, Y. Shadmi and Y. Soreq for valuable comments and discussions, and for reviewing this work.
We thank Y. Cohen, Y. Weiss, G. Durieux, T. Kitahara and G. Sierra for useful discussions.
This research was supported by a grant from the United States-Israel Binational Science Foundation (BSF), Jerusalem, Israel, by a grant from the Israel Science Foundation (ISF), and by grant FIS2017-84368-P from Spain's MINECO. 

\appendix

\section{Analysis of the entanglement criteria}\label{app:criteria}

We derive here some useful results related to the entanglement criteria discussed in Sec.~\ref{subsec:Hilbert}. In order to further analyze the Peres-Horodecki criterion, we expand the general expression of Eq. (\ref{eq:GeneralBipartiteStateRotations}) for a density matrix in a Hilbert space of $2\times 2$ dimensions:
\begin{widetext}
\begin{equation}\label{eq:GeneralBipartiteStateExplicitYoav}
\rho=\frac{1}{4}\left[\begin{array}{cccc}
1+B^{+}_3+B^{-}_{3}+C_{33} & B^{-}_1+C_{31}-i(B^{-}_2+C_{32}) & B^{+}_1+C_{13}-i(B^{+}_2+C_{23}) & C_{11}-C_{22}-i(C_{12}+C_{21}) \\
B^{-}_1+C_{31}+i(B^{-}_2+C_{32})& 1+B^{+}_3-B^{-}_{3}-C_{33} & C_{11}+C_{22}+i(C_{12}-C_{21}) & B^{+}_1-C_{13}-i(B^{+}_2-C_{23}) \\
B^{+}_1+C_{13}+i(B^{+}_{2}+C_{23}) & C_{11}+C_{22}+i(C_{21}-C_{12}) & 1-B^{+}_3+B^{-}_{3}-C_{33} & B^{-}_1-C_{31}-i(B^{-}_{2}-C_{32})\\
C_{11}-C_{22}+i(C_{21}+C_{12}) & B^{+}_1-C_{13}+i(B^{+}_2-C_{23}) & B^{-}_{1}-C_{31}+i(B^{-}_{2}-C_{32}) & 1-B^{+}_3-B^{-}_{3}+C_{33}\\
\end{array}\right]
\end{equation}
where we are working in the spin basis along the third spin component, $\ket{\uparrow\uparrow},\ket{\uparrow\downarrow},\ket{\downarrow\uparrow},\ket{\downarrow\downarrow}$. Taking the partial transpose of $\rho$ with respect to the second subsystem, which amounts to transpose the $4$ blocks of size $2\times 2$ of $\rho$, gives

\begin{equation}\label{eq:GeneralBipartiteStateExplicitYoavTranspose}
\rho^{\rm{T_2}}=\frac{1}{4}\left[\begin{array}{cccc}
1+B^{+}_3+B^{-}_{3}+C_{33} & B^{-}_1+C_{31}+i(B^{-}_2+C_{32}) & B^{+}_1+C_{13}-i(B^{+}_2+C_{23}) & C_{11}+C_{22}+i(C_{12}-C_{21}) \\
B^{-}_1+C_{31}-i(B^{-}_2+C_{32})& 1+B^{+}_3-B^{-}_{3}-C_{33} & C_{11}-C_{22}-i(C_{12}+C_{21}) & B^{+}_1-C_{13}-i(B^{+}_2-C_{23}) \\
B^{+}_1+C_{13}+i(B^{+}_{2}+C_{23}) & C_{11}-C_{22}+i(C_{21}+C_{12}) & 1-B^{+}_3+B^{-}_{3}-C_{33} & B^{-}_{1}-C_{31}+i(B^{-}_{2}-C_{32}) \\
C_{11}+C_{22}+i(C_{21}-C_{12}) & B^{+}_1-C_{13}+i(B^{+}_2-C_{23}) & B^{-}_1-C_{31}-i(B^{-}_{2}-C_{32}) & 1-B^{+}_3-B^{-}_{3}+C_{33}\\
\end{array}\right]
\end{equation}

The Peres-Horodecki criterion asserts that $\rho^{\rm{T_2}}$ is nonnegative if and only if $\rho$ is separable. In particular, if $\rho^{\rm{T_2}}$ is nonnegative, the form arising from considering vectors with only first and fourth component, i.e.,
\begin{equation}\label{eq:GeneralBipartiteStateExplicitYoavTranspose}
\rho_{C}\equiv\left[\begin{array}{cc}
1+B^{+}_3+B^{-}_{3}+C_{33} & C_{11}+C_{22}+i(C_{12}-C_{21}) \\
C_{11}+C_{22}+i(C_{21}-C_{12}) & 1-B^{+}_3-B^{-}_{3}+C_{33}\\
\end{array}\right]
\end{equation}
\end{widetext}
must be also nonnegative, which implies $\det \rho_{C}\geq 0$. Hence, $\det \rho_{C}<0$ is a \textit{sufficient} condition for the presence of entanglement, which can be written as $\mathcal{P}>0$, with
\begin{equation}
\mathcal{P}\equiv(B^{+}_3+B^{-}_3)^2+(C_{11}+C_{22})^2+(C_{21}-C_{12})^2-(1+C_{33})^2
\end{equation}
Furthermore, since
\begin{equation}
\mathcal{P}\geq (C_{11}+C_{22})^2-(1+C_{33})^2\equiv \tilde{\mathcal{P}},
\end{equation}
$\tilde{\mathcal{P}}>0$ is by itself a sufficient condition for entanglement. Specifically, as $1+C_{33}\geq0$, $\tilde{\mathcal{P}}>0$ if and only if
\begin{equation}
|C_{11}+C_{22}|>1+C_{33}
\end{equation}
Thus,
\begin{equation}\label{eq:DeltaGeneralized}
\Delta\equiv-C_{33}+|C_{11}+C_{22}|-1>0
\end{equation}
is a sufficient condition for entanglement, which in particular implies that
\begin{equation}\label{eq:DeltaTraceGeneralized}
-\textrm{tr}[\mathbf{C}]>1
\end{equation}
is also a sufficient condition for entanglement. We remark that all the above derived criteria are general sufficient conditions for entanglement, valid for arbitrary quantum states.

We can further obtain analytical results by making some assumptions on the form of $\rho$. For instance, we consider unpolarized quantum states ($B^{+}_i=B^{-}_i=0$) whose correlation matrix is symmetric ($C_{ij}=C_{ji}$), as it is the case of the density matrix $\rho(M_{t\bar{t}},\hat{k})$ describing $t\bar{t}$ production at LO (see Sec.~\ref{sec:entanglement}). In that situation, we can diagonalize the correlation matrix by the appropriated rotation, so $C=\textrm{diag}[C_1,C_2,C_3]$ and thus the density matrix of Eq. (\ref{eq:GeneralBipartiteStateExplicitYoav}) is reduced to
\begin{equation}\label{eq:GeneralBipartiteStateExplicitYoavDiagonal}
\rho=\frac{1}{4}\left[\begin{array}{cccc}
1+C_{3} & 0 & 0 & C_{1}-C_{2} \\
0 & 1-C_{3} & C_{1}+C_{2} & 0 \\
0  & C_{1}+C_{2} & 1-C_{3} & 0\\
C_{1}-C_{2} & 0 & 0 & 1+C_{3}\\
\end{array}\right]
\end{equation}
It is easy to see that
\begin{equation}\label{eq:PhysicalityRho}
\pm C_{3}+|C_{1}\pm C_{2}|-1\leq 0
\end{equation}
by demanding $\rho$ to be a physical state, while the Peres-Horodecki criterion implies that the state is entangled whenever
\begin{equation}\label{eq:EntanglementPeresHorodeckiDiagonal}
\pm C_{3}+|C_{1}\mp C_{2}|-1>0
\end{equation}
In particular, the $-+$ combination in the above equation is the condition $\Delta>0$ in terms of the correlations $C_i$. We note that, by combining Eq. (\ref{eq:PhysicalityRho}) with Eq. (\ref{eq:EntanglementPeresHorodeckiDiagonal}), only the violation where $\pm C_3>0$ can be achieved. Hence, the state is entangled if and only if
\begin{eqnarray}\label{eq:EntanglementPeresHorodeckiDiagonalSign}
-C_3+|C_1+C_2|-1&>&0,~C_3\leq 0\\
\nonumber C_3+|C_1-C_2|-1&>&0,~C_3\geq 0
\end{eqnarray}

In the particular case of Eq. (\ref{eq:GeneralBipartiteStateExplicitYoavDiagonal}) where $\rho$ is real and unpolarized, the concurrence can be also analytically computed from its definition, Eq. (\ref{eq:Concurrence}). Since $\tilde{\rho}=\rho$ if $\rho$ is real and unpolarized, $\mathcal{C}(\rho)=\rho$, and then we only need to obtain the eigenvalues of $\rho$, which are $\frac{1}{4}(1+C_3\pm|C_{1}-C_{2}|)$,~$\frac{1}{4}(1-C_3\pm|C_{1}+C_{2}|)$. Moreover, since
\begin{equation}
\sum_i \lambda_i=1,
\end{equation}
the concurrence is just $C[\rho]=\max(2\lambda_1-1,0)$, with $2\lambda_1-1=(\pm C_{3}+|C_{1}\mp C_{2}|-1)/2$. By noting the analogy with Eq. (\ref{eq:EntanglementPeresHorodeckiDiagonal}), we finally arrive at
\begin{eqnarray}\label{eq:ConcurrenceSign}
\nonumber C[\rho]&=&\frac{1}{2}\max[-C_3+|C_1+C_2|-1,0],~C_3\leq 0\\
 C[\rho]&=&\frac{1}{2}\max[C_3+|C_1-C_2|-1,0],~C_3\geq 0
\end{eqnarray}
which is completely equivalent to the entanglement criterion in Eq. (\ref{eq:EntanglementPeresHorodeckiDiagonalSign}).

Another useful example is provided by the particular case in which there is invariance under rotations along a certain axis, which can be chosen without loss of generality along the third component of the spin. This is the situation of the quantum state of the $t\bar{t}$ pair after averaging over all top directions (see Sec.~\ref{sec:totalquantumstate}). In that case, the spin polarizations must be along the symmetry $z$-axis, $B^{\pm}_{i}=B^{\pm}_{z}\delta_{i3}$, and the correlation matrix is diagonal, $C_{ij}=\delta_{ij}C_j$, with eigenvalues $C_{1}=C_{2}=C_{\perp}$ and $C_{3}=C_{z}$. The Peres-Horodecki criterion is then equivalent to the condition
\begin{equation}\label{eq:PeresHorodeckiAxial}
4C^2_{\perp}+(B^{+}_{z}+B^{-}_{z})^2-(1+C_z)^2>0
\end{equation}
In particular, for an unpolarized quantum state,
\begin{equation}\label{eq:PeresHorodeckiAxialLO}
\delta\equiv-C_z+2|C_{\perp}|-1>0
\end{equation}
is a necessary and sufficient condition for entanglement, while for $B^{\pm}_z\neq 0$, $\delta>0$ is just a sufficient condition.

\section{Leading-order calculation of the spin density matrix}\label{app:LODensityMatrix}

We detail here the computation of the spin density matrix of the $t\bar{t}$ pair at LO. For that purpose, we review the main results about spin correlations of $t\bar{t}$ pairs, which can be found in the usual literature on the topic~\cite{Bernreuther1994,Bernreuther1998,Bernreuther2004,Uwer2005,Baumgart2013,Bernreuther2015} and are needed for the computation of $\rho(M_{t\bar{t}},\hat{k})$.
In particular, as mentioned in the main text, at LO only $5$ parameters are needed to characterize the production spin density matrix in the helicity basis: $\tilde{A},\tilde{C}_{kk},\tilde{C}_{nn},\tilde{C}_{rr},\tilde{C}_{kr}$. The values of these coefficients for the partonic production spin density matrices $R^{I}$ are well known~\cite{Bernreuther1994,Bernreuther1998,Uwer2005}. Specifically, for $I=q\bar{q}$,
\begin{eqnarray}\label{eq:LOSpinCorrelationsqq}
\tilde{A}^{q\bar{q}}&=&F_q(2-\beta^2\sin^2\Theta)\\
\nonumber \tilde{C}^{q\bar{q}}_{rr}&=&F_q(2-\beta^2)\sin^2\Theta\\
\nonumber \tilde{C}^{q\bar{q}}_{nn}&=&-F_q\beta^2\sin^2\Theta\\
\nonumber \tilde{C}^{q\bar{q}}_{kk}&=&F_q(2\cos^2\Theta+\beta^2\sin^2\Theta)\\
\nonumber \tilde{C}^{q\bar{q}}_{rk}&=&\tilde{C}^{q\bar{q}}_{kr}=F_q\sqrt{1-\beta^2}\sin2\Theta
\end{eqnarray}
while, for $I=gg$,
\begin{eqnarray}\label{eq:LOSpinCorrelationsgg}
\tilde{A}^{gg}&=&F_g(\Theta)\left[1+2\beta^2\sin^2\Theta-\beta^4(1+\sin^4\Theta)\right]\\
\nonumber \tilde{C}^{gg}_{rr}&=&-F_g(\Theta)\left[1-\beta^2(2-\beta^2)(1+\sin^4\Theta)\right]\\
\nonumber \tilde{C}^{gg}_{nn}&=&-F_g(\Theta)\left[1-2\beta^2+\beta^4(1+\sin^4\Theta)\right]\\
\nonumber \tilde{C}^{gg}_{kk}&=&-F_g(\Theta)\left[1-\beta^2\frac{\sin^2 2\Theta}{2}-\beta^4(1+\sin^4\Theta)\right]\\
\nonumber \tilde{C}^{gg}_{rk}&=&F_g(\Theta)\sqrt{1-\beta^2}\beta^2\sin2\Theta\sin^2\Theta
\end{eqnarray}
where the normalization factors $F_q,F_g(\Theta)$ are
\begin{equation}
    F_q=\frac{1}{18},~F_g(\Theta)=\frac{7+9\beta^2\cos^2\Theta}{192(1-\beta^2\cos^2\Theta)^2}
\end{equation}

The production spin density matrix $R$ is computed in terms of the matrices $R^{I}$ as in Eq.~(\ref{eq:Rtotal}), using the computed luminosity functions $L^{I}(M_{t\bar{t}})$.
Finally, the actual spin density matrices $\rho^{I},\rho$ are obtained by normalizing $R^{I},R$ to have unit trace, respectively.

\section{Angular averaging}\label{app:AngularAveraging}

We compute here the distributions and correlations resulting from averaging over the angular coordinates. We first characterize the angular-averaged production spin density matrix
\begin{eqnarray}\label{eq:CorrelationAzimuthAveraged}
R_{\Omega}(M_{t\bar{t}})&=&\frac{1}{4\pi}\int\mathrm{d}\Omega~R(M_{t\bar{t}},\hat{k})
\end{eqnarray}
Specifically, we begin by computing the partonic matrices $R^I_{\Omega}(M_{t\bar{t}})$ since they can be analytically found from Eqs. (\ref{eq:LOSpinCorrelationsqq}), (\ref{eq:LOSpinCorrelationsgg}). They are characterized by $3$ parameters: $\tilde{A}^I(M_{t\bar{t}}),\tilde{C}^I_{\perp}(M_{t\bar{t}})$ and $\tilde{C}^I_z(M_{t\bar{t}})$, $\tilde{A}^I(M_{t\bar{t}})$ being simply the angular average of $\tilde{A}^I(M_{t\bar{t}},\hat{k})$.

Regarding the computation of the spin correlation matrix, after averaging over the azimuth, we are left with a diagonal matrix in the $\{\hat{x},\hat{y},\hat{z}\}$ beam basis. The polar integrals in $\Theta$ for $I=q\bar{q}$ are polynomials in $t=\cos\Theta$, easily solvable as $\mathrm{d}\Theta\sin\Theta=\mathrm{d}t$; in terms of $\sin\Theta$, we deal with integrals of the form
\begin{equation}
F_n\equiv\int^{\frac{\pi}{2}}_0\mathrm{d}\Theta~\sin^n\Theta=\frac{\Gamma\left(\frac{n+1}{2}\right)}{\Gamma\left(\frac{n+2}{2}\right)}\frac{\sqrt{\pi}}{2}
\end{equation}
from where we find
\begin{eqnarray}\label{eq:AngularAveragedCorrelationsqq}
\tilde{A}^{q\bar{q}}(M_{t\bar{t}})&=&\frac{1}{9}\left[1-\frac{\beta^2}{3}\right]\\
\nonumber C_{\perp}^{q\bar{q}}(M_{t\bar{t}})&=&\frac{2}{135}f(\beta)\\
\nonumber C_{z}^{q\bar{q}}(M_{t\bar{t}})&=&\frac{1}{9}\left[1-\frac{\beta^2}{3}-\frac{4}{15}f(\beta)\right]\\
\nonumber f(\beta)&\equiv&\frac{\left(1-\sqrt{1-\beta^2}\right)^2}{2}
\end{eqnarray}

On the other hand, the angular averages for $gg$ processes are a little more involved. It is seen that all the appearing integrals can be put in terms of $K_{n,2}(\beta)$, with
\begin{equation}
K_{n,m}(x)\equiv\int^{x}_{-x}\mathrm{d}z~\frac{z^{2n}}{(1-z^2)^m}
\end{equation}
The above integral satisfies the following recursion relations
\begin{widetext}
\begin{eqnarray}
 K_{n,m}(x)&=&K_{n-1,m}(x)-K_{n-1,m-1}(x),~K_{n,0}(x)=2\frac{x^{2n+1}}{2n+1}\\
\nonumber K_{0,m}(x)&=&\frac{1}{2(m-1)}\left[\frac{2x}{(1-x^2)^{m-1}}+(2m-3)K_{0,m-1}(x)\right],~m>1 \\
\nonumber K_{0,1}(x)&=&2\textrm{atanh}(x)
\end{eqnarray}
from which we arrive at
\begin{eqnarray}
K_{n,1}(x)&=&K_{n-1,1}(x)-K_{n-1,0}(x)=2\left[\textrm{atanh}(x)-\sum^{n-1}_{k=0}\frac{x^{2k+1}}{2k+1}\right]\\
\nonumber K_{n,2}(x)&=&K_{n-1,2}(x)-K_{n-1,1}(x)=\frac{x}{1-x^2}-(2n-1)\textrm{atanh}(x)+\sum^{n-2}_{k=0}\frac{2(n-1-k)}{2k+1}x^{2k+1}
\end{eqnarray}
With the help of these results, we find that
\begin{eqnarray}\label{eq:AngularAveragedCorrelationsgg}
\tilde{A}^{gg}(M_{t\bar{t}})&=&
\frac{1}{192}\left[-59+31\beta^2+(66-36\beta^2+2\beta^4)\frac{\textrm{atanh}(\beta)}{\beta}\right]\\
\nonumber C_{\perp}^{gg}(M_{t\bar{t}})&=&
\frac{1-\beta^2}{192}\left[9-16\frac{\textrm{atanh}(\beta)}{\beta}\right]+g(\beta)\\
\nonumber C_{z}^{gg}(M_{t\bar{t}})&=&\frac{1}{192}\left[-109+49\beta^2+(102-72\beta^2+2\beta^4)\frac{\textrm{atanh}(\beta)}{\beta}\right]-2g(\beta)\\
\nonumber g(\beta)&\equiv&\frac{f(\beta)}{96\beta^4}\left[49-\frac{149}{3}\beta^2+\frac{24}{5}\beta^4-(17\beta^4-66\beta^2+49)\frac{\textrm{atanh}(\beta)}{\beta}\right]
\end{eqnarray}

Similar considerations arise when computing the angular average of the expectation value of the spin correlations in the helicity basis [see Eq. (\ref{eq:Expectationvalue})], obtaining for $I=q\bar{q}$
\begin{eqnarray}\label{eq:AngularAveragedCorrelationsHelicityqq}
\nonumber \tilde{C}^{q\bar{q}}_{rr}(M_{t\bar{t}})&=&\frac{(2-\beta^2)}{27}\\
\tilde{C}^{q\bar{q}}_{nn}(M_{t\bar{t}})&=&-\frac{\beta^2}{27}\\
\nonumber \tilde{C}^{q\bar{q}}_{kk}(M_{t\bar{t}})&=&\frac{1+\beta^2}{27}
\end{eqnarray}
and for $I=gg$
\begin{eqnarray}\label{eq:AngularAveragedCorrelationsHelicitygg}
\nonumber \tilde{C}^{gg}_{rr}(M_{t\bar{t}})&=&-\frac{1}{192}
\left[87-31\beta^2+66\frac{\frac{\textrm{atanh}(\beta)}{\beta}-1}{\beta^2}-
\left(102-38\beta^2+2\beta^4\right)\frac{\textrm{atanh}(\beta)}{\beta}\right]\\
\tilde{C}^{gg}_{nn}(M_{t\bar{t}})&=&-\frac{1}{192}
\left[41-31\beta^2-\left(34-36\beta^2+2\beta^4\right)\frac{\textrm{atanh}(\beta)}{\beta}\right]\\
\nonumber \tilde{C}^{gg}_{kk}(M_{t\bar{t}})&=&-\frac{1}{192}
\left[-37+31\beta^2-66\frac{\frac{\textrm{atanh}(\beta)}{\beta}-1}{\beta^2}+
\left(66-34\beta^2+2\beta^4\right)\frac{\textrm{atanh}(\beta)}{\beta}\right]
\end{eqnarray}
\end{widetext}

We note that the resulting matrix $\tilde{C}^I_{ij}(M_{t\bar{t}})$ in the helicity basis is also diagonal after the angular averaging since $\tilde{C}^I_{kr}(M_{t\bar{t}})$ vanishes at LO.

The total angular-averaged production spin density matrix $R_{\Omega}(M_{t\bar{t}})$ is computed in terms of its partonic counterparts from Eq. (\ref{eq:Rtotal}). Finally, the associated quantum state and spin correlations will be simply given by
\begin{equation}
\rho_{\Omega}(M_{t\bar{t}})=\frac{R_{\Omega}(M_{t\bar{t}})}{4\tilde{A}(M_{t\bar{t}})},~C_{ij}(M_{t\bar{t}})=\frac{\tilde{C}_{ij}(M_{t\bar{t}})}{4\tilde{A}(M_{t\bar{t}})}
\end{equation}
\bibliographystyle{apsrev4-1}
\bibliography{Entanglement.bib}

\end{document}